\documentclass[12pt]{article}

\pdfoutput=1

\usepackage{latexsym}
\usepackage{epsfig,amssymb,euscript}
\usepackage{amsmath}
\usepackage{mathrsfs}
\usepackage{ccaption}    
\usepackage[usenames,dvipsnames]{color}
\usepackage{subfigure}

\usepackage[
      colorlinks=true,
      linkcolor=blue,
      urlcolor=blue,
      filecolor=black,
      citecolor=red,
      pdfstartview=FitV,
      pdftitle={},
        pdfauthor={Simon A. Gentle, Mukund Rangamani},
        pdfsubject={},
        pdfkeywords={},
        pdfpagemode={},
        bookmarksopen=true
      ]{hyperref}

\makeatletter
\@addtoreset{equation}{section}

\makeatletter
\renewcommand\section{\@startsection {section}{1}{\z@}%
                                   {-3.5ex \@plus -1ex \@minus -.2ex}
                                   {2.3ex \@plus.2ex}%
                                   {\normalfont\large\bfseries}}
\renewcommand\subsection{\@startsection{subsection}{2}{\z@}%
                                     {-3.25ex\@plus -1ex \@minus -.2ex}%
                                     {1.5ex \@plus .2ex}%
                                     {\normalfont\bfseries}}

  \captionnamefont{\bfseries}
  \captiontitlefont{\small\sffamily}
  \captiondelim{: }
  \hangcaption

\usepackage{ dsfont } 

\definecolor{rust}{rgb}{0.8,0.2,0.2}
\definecolor{green}{rgb}{0.1,0.8,0.2}
\definecolor{hue55}{rgb}{0.0,0.5,0.67}
\definecolor{hue75}{rgb}{0.65,0.15,0.6}
\definecolor{light-gray}{gray}{0.5}

\def\AdS#1{AdS$_{#1}$}

\def\vev#1{\langle #1 \rangle}

\def\cwedge{\blacklozenge_{{\cal A}}} 
\def\domd{\lozenge_{\cal A}}  
\def\bcwedge{\partial_{{\cal M}}(\blacklozenge_{{\cal A}})}
\def\bcwedgef{\partial_+(\blacklozenge_{{\cal A}})}
\def\bcwedgep{\partial_-(\blacklozenge_{{\cal A}})}
\def\entsurf{\partial {\cal A}} 
\def\extr#1{{\mathfrak E}_{#1}}   
\def\csf#1{{\Xi}_{#1}}  
\def\rE{r_{{\mathfrak E}}}   
\def\rX{r_{\Xi}}  

\title{\bf Holographic entanglement and causal information in coherent states}

\author{\large Simon A.~Gentle$^a$ and Mukund Rangamani$^b$\\ \\
        \small $^a$\it Department of Physics and Astronomy, \\
        \small \it University of California, Los Angeles, CA 90095, USA \\ \\
        \small $^b$\it Centre for Particle Theory \& Department of Mathematical Sciences, \\
        \small \it Lower Mountjoy, South Road, Durham DH1 3LE, UK \\ \\
       \normalsize\href{mailto:sgentle@physics.ucla.edu}{\texttt{sgentle@physics.ucla.edu}}\texttt{, }\href{mailto:mukund.rangamani@durham.ac.uk}{\texttt{mukund.rangamani@durham.ac.uk}}}
\date{}

\begin{document}

\setlength{\baselineskip}{18pt}

\maketitle

\begin{picture}(0,0)(0,0)
\put(350,300){DCPT-13/37}
\end{picture}
\vspace{-36pt}

\thispagestyle{empty}                        

\begin{abstract}
Scalar solitons in global AdS$_4$ are holographically dual to coherent states carrying a non-trivial condensate of a scalar operator.  We  study the holographic information content of these states, focusing on a particular spatial region, by examining the entanglement entropy and  causal holographic information. We show  generically that whenever the dimension of the condensed operator is sufficiently low (characterized by the double-trace operator becoming relevant), such coherent states have lower entanglement and causal holographic information than the vacuum state of the system, despite  having greater energy. We also use these geometries to illustrate the fact that causal wedges associated with a simply-connected boundary region can have non-trivial topology even in causally trivial spacetimes. 
\end{abstract}

\pagebreak
\setcounter{page}{1}

\tableofcontents

\section{Introduction}
\label{sec:intro}

The gravitational description of strongly-coupled large $N$ quantum field theories provides an important perspective on their quantum dynamics via the AdS/CFT correspondence. In recent years we have seen the correspondence provide examples of gravitational solutions that show striking similarities with real-world physics. In particular, many phenomena that we encounter in quantum many-body systems have very simple and elegant incarnations in the gravitational setting.

One natural question in quantum many-body systems is the nature of entanglement in a given quantum state. In a certain sense this quantity encodes the information about the correlations between the microscopic quanta that build up the state. In the holographic context this notion takes on an even more important meaning since one can associate the quantum entanglement of the state with elements of the dual spacetime geometry --- as evidenced by the conjectures of \cite{VanRaamsdonk:2009ar,VanRaamsdonk:2010pw,Maldacena:2013xja}. The basic premise behind these ideas is the geometrization of quantum entanglement in the holographic context by the Ryu-Takayanagi prescription \cite{Ryu:2006bv, Ryu:2006ef} and its covariant generalization \cite{Hubeny:2007xt}.\footnote{ The minimal surface prescription of   \cite{Ryu:2006bv, Ryu:2006ef} was recently established on firm footing by the analysis of \cite{Lewkowycz:2013nqa}.} These ideas provide an avenue to describe how spacetime geometry can emerge from the underlying quantum mechanical description.
 
Despite these various fascinating developments, to date a clear understanding of how boundary field theories can holographically reconstruct spacetime remains elusive. Various attempts to address this question have of course been undertaken over the years, ranging from using the WKB approximation of correlators in terms of geodesics to detect spacetime structures \cite{Balasubramanian:1999zv, Louko:2000tp,Gregory:2000an,Kraus:2002iv, Fidkowski:2003nf, Hubeny:2006yu}, to using this data and entanglement entropy to reconstruct spacetime geometry \cite{Hammersley:2006cp,Hammersley:2007ab,Bilson:2008ab,Bilson:2010ff,Balasubramanian:2013lsa}. A summary of the early developments and a critical discussion of some of the limitations of various methods can be found in \cite{Hubeny:2010ry, Hubeny:2012ry}.

In the recent past the focus has been on understanding how local regions of field theory can lead to spacetime reconstruction, in an attempt to try to distill the information content of spacetime geometry into reduced density matrices of the field theory \cite{Bousso:2012sj, Czech:2012bh, Hubeny:2012wa, Bousso:2012mh}. Emerging from these discussions is the central role played by regions in the bulk associated with two distinct constructs: (i) extremal surfaces that capture entanglement entropy \cite{Ryu:2006bv,Hubeny:2007xt} and (ii) causal information surfaces built from causal domains associated with a given boundary region \cite{Bousso:2012sj, Hubeny:2012wa}.
It has been argued in \cite{Hubeny:2012wa} that the area of the causal information surface $\csf{\cal A} $ associated with a boundary region ${\cal A}$ captures the minimal amount of information contained in the region relevant for the holographic reconstruction (see \cite{Freivogel:2013zta, Kelly:2013aja} for proposals to interpret this quantity in field theory). On the other hand, a study of the extremal surface $\extr{\cal A}$ (whose area computes the entanglement entropy) suggests that the natural bulk region  associated with the reduced density matrix $\rho_{\cal A}$ of the boundary region  is larger. It is roughly given by the bulk domain of dependence of a spacelike region bounded by ${\cal A}$ on the boundary and $\extr{\cal A} $ in the bulk \cite{Czech:2012bh}. 

While the abstract concepts are interesting in their own right, recent explorations of the causally motivated constructions have revealed some interesting surprises and a curious interplay between the surfaces $\csf{\cal A}$ and $\extr{\cal A}$. Firstly, it has been shown on very general grounds that the extremal surface lies outside the causal information  surface \cite{Hubeny:2012wa,Wall:2012uf, Hubeny:2013gba}.\footnote{ The result of \cite{Wall:2012uf} (Theorem 6) is actually stronger and says that the extremal surface $\extr{\cal A}$ is spacelike separated from $\csf{\cal A}$. For the geometries we consider, this will trivially be true, since both surfaces lie on the same time-slice.} More curiously, the bulk causal wedge associated to a simply-connected  region ${\cal A}$ on the boundary can itself have non-trivial topology \cite{Hubeny:2013gba}. These two observations in turn lead  to non-trivial constraints on entanglement entropy, such as the saturation of the Araki-Lieb inequality for finite systems in a density matrix \cite{Hubeny:2013gta}.

While these results are interesting, one should bear in mind that they have been explicitly obtained in sufficiently simple states of the field theory. Indeed, much of the discussion hitherto has been restricted to either the vacuum of the field theory (hence the AdS spacetime) or a thermal density matrix (the Schwarzschild-AdS black hole geometry). Clearly these are special states in the field theory and one would like to have some intuition of how the measures of quantum information encoded in the entanglement entropy and the causal holographic information behave in other geometries. 

In this paper we therefore  explore these ideas in a further simple class of spacetimes: charged scalar solitons in global AdS$_4$. These geometries are static, spherically-symmetric solutions to Einstein-Maxwell-scalar systems and have been studied previously in various contexts. Perturbative constructions of these solutions (and also black hole excitations about them) were first described in \cite{Basu:2010uz, Bhattacharyya:2010yg} (in AdS$_5$). A more detailed analysis of such geometries was undertaken in \cite{Gentle:2011kv,Dias:2011tj}  (see also \cite{Hartmann:2013kna, Kichakova:2013sza} for some recent developments). The phase space of these solutions was shown to be extremely rich,  with multiple branches of solitonic solutions  depending on the quantum numbers  (dimension and  global conserved charge) of the scalar operator ${\cal O}$. We focus on the analysis of \cite{Gentle:2011kv} carried out in AdS$_4$  for  simple phenomenological models and also top-down models coming from consistent truncations of eleven-dimensional supergravity. One can view the multitude of solutions as arising due to a competition between the charge repulsion and the gravitational attraction. 
 
 From a field theory perspective these solitons are atypical states in the microcanonical ensemble with energies (or conformal dimensions) and charges\footnote{ We focus on states with bulk energies of order $O(N^\frac{3}{2})$ in \AdS{4} as these are geometries where the gravitational backreaction of the matter fields is $O(1)$.}  $E, Q\sim N^\frac{3}{2}$. Typical states of this ensemble would of course be the black hole micro-states. Not only are these states atypical from the statistical perspective, they are also curious from another viewpoint. While small perturbations about global AdS tend to collapse into a black hole rather quickly \cite{Bizon:2011gg}, there are analytical arguments \cite{Dias:2012tq} and numerical evidence based on the study of time-periodic solutions \cite{Maliborski:2013jca} and boson stars \cite{Buchel:2013uba} that this rapid collapse is a prerogative of vacuum AdS spacetime and that excited states are  immune from such behaviour in general. This is rather bizarre from the field theory perspective: since the collapse process maps to thermalization in the field theory, it suggests that there are states of a large $N$ field theory that do not thermalize upon being perturbed.  While there is no explicit evidence that the charged scalar solitons described above are of this type, they are sufficiently similar to the boson star geometries mentioned above that one would suspect they too are immune from rapid thermalization.

These observations make the soliton states quite interesting from a microscopic perspective. As a result it would be useful to know how these pure quantum states behave in their holographic information content.\footnote{ A similar analysis for a class of related boson star geometries was undertaken in \cite{Nogueira:2013if}.} In this paper we  undertake this exercise and uncover some interesting properties of these coherent states. To keep things simple we will consider regions in the field theory that preserve a $U(1) \subset SO(3)$ symmetry. The field theory region ${\cal A}$ whose information content we explore will be a polar-cap of the boundary ${\bf S}^2$. 

Our analysis reveals one rather surprising result: for coherent states built from a macroscopic population of bosonic modes dual to relevant operators that allow multi-trace deformation (relevant or marginal) in the field theory, the entanglement entropy measured relative to the vacuum is negative! To be specific, if we consider field theory operators ${\cal O}_\Delta$ with $\Delta = \Delta_-$ obtained by imposing Neumann (or alternate) boundary conditions on the scalar field in the bulk, then we find that the solitons built from these operators have lower entanglement than the vacuum. This is in marked contrast to the solitons where we impose Dirichlet (or standard) quantization $\Delta = \Delta_+$ --- these have positive entanglement relative to the vacuum.\footnote{ A qualitative difference between the two quantizations in the behaviour of the entanglement entropy was observed for holographic superconductors in \cite{Albash:2012pd}.} We should note that this is despite the fact that the solutions carry positive energy relative to the vacuum and is consistent with the observations made based on studies of relative entropy in \cite{Blanco:2013joa}. This observation poses some interesting challenges for generalizing the map between entanglement and linearized Einstein's equations that exploits the linear relation between the entanglement relative to the vacuum and the change in the modular Hamiltonian ($\Delta S = \Delta H$) \cite{Blanco:2013joa, Lashkari:2013koa} (see also \cite{Nozaki:2013vta, Bhattacharya:2013bna} for a somewhat different take on this issue).
This reduction is also seen in the case of the causal holographic information. 

Our study of causal wedges in these soliton geometries also provides an explicit example of a causally trivial spacetime that has a causal wedge with non-trivial topology (for simply-connected polar-cap regions). The geometries where we find this behaviour also happen to admit null circular orbits (around the core of the soliton). This is  consistent with the general conjecture of \cite{Hubeny:2013gba} who had earlier demonstrated non-trivial causal wedge topology explicitly in a black hole spacetime and gave arguments for why the phenomenon should persist even for causally trivial geometries. As a natural by-product of our analysis we will be able to gain some insight into the behaviour of bulk-cone singularities \cite{Hubeny:2006yu} in these states. 

This paper is organised as follows.  We will begin with an overview of some background material: in \S\ref{sec:solitons} we describe the class of solitonic solutions we study and then summarize the basic definitions of the holographic measures of information in \S\ref{sec:holoinfo}.  We present our main results for the entanglement entropy in \S\ref{sec:ee}. We then turn to the causal construction and describe the salient results in \S\ref{sec:wedges}. We conclude with a discussion in \S\ref{sec:discussion}.

\section{Coherent states in the CFT and information measures}
\label{sec:review}
To set the stage for our discussion we review some of the salient properties of the objects we are interested in. Firstly, in \S\ref{sec:solitons} we introduce the class of CFT states we focus on, describing them in terms of their dual geometry as charged scalar solitons in global \AdS{4}. We then quickly summarize the necessary details of the observables we will study in these states using holographic methods in \S\ref{sec:holoinfo}.

\subsection{Scalar solitons in global AdS}
\label{sec:solitons}%

The coherent (pure) states of the CFT we are interested in are condensates of bosonic modes and can be described in terms of scalar solitons that are asymptotically globally \AdS{4}.  The family of solitons we study can be described in the bulk \AdS{4} in an Einstein-Maxwell-scalar theory
\begin{equation}\label{eq:generalaction}
S = \frac{1}{16\pi G_4} \int d^4x \sqrt{-g} \left(R  - \frac{1}{4}F^2 - \left(\partial\phi\right)^2 - \frac{1}{L^2}\, Q(\phi) A^2 -\frac{1}{L^2}\, V(\phi)\right)
\end{equation}
with $F=dA$.  We have written the action in a gauged-fixed form: $\phi$ is the norm of a complex scalar and we have chosen to absorb the phase into the gauge field. Complete specification of the bulk theory requires details of the functions $\{V(\phi), Q(\phi)\}$. In \cite{Gentle:2011kv} three distinct examples were considered and we focus on two of these here:
\begin{align}
\text{Phenomenological model}: & \quad V(\phi) = -6 -2\,\phi^2 , \quad Q(\phi) = q^2\phi^2 
\label{eq:phenmodel} \\
U(1)^4\;\; \text{truncation}:& \quad V(\phi)=  -2\,\left(2+\cosh{\sqrt{2}\phi}\right), \quad Q(\phi) = \frac{1}{2} \,\sinh^2\frac{\phi}{\sqrt{2}}
\label{eq:DGmodel}
\end{align}
Here, $L$ is the \AdS{4} radius, with $L^2/G_4 \sim N^\frac{3}{2}$. Note that in both examples the scalar mass is $m_\phi^2\, L^2 = -2$, so that by the standard AdS/CFT dictionary we have a binary choice for the dual CFT. We either have a boundary operator ${\cal O}_2$ of dimension $\Delta = 2$ (standard or Dirichlet boundary condition) or a boundary operator ${\cal O}_1$ of dimension $\Delta = 1$ (alternate or Neumann boundary condition). In the phenomenological theory, $q$ is a free parameter that we can vary.  From now on we set $L=1$.\footnote{ In these units the 4-dimensional Newton's constant is related to the effective central charge $c_\text{eff}$ of the dual CFT via $c_\text{eff} = \left(16\pi\, G_4\right)^{-1}$. For the M2-brane world-volume theory we have $c_\text{eff} = \frac{1}{48\pi}\,(2N)^{3/2}$.}

We are interested in static, spherically-symmetric solutions that preserve the $\mathbb{R}\times SO(3)$  and asymptote to global AdS$_4$. The metric ansatz is 
\begin{equation}\label{eq:metricansatz}
ds^2 = - g(r)e^{-\beta(r)} dt^2+ \frac{dr^2}{g(r)}  + r^2 \left(d\theta^2+\sin^2\theta\, d\varphi^2\right)\,,
\end{equation}
while for the vector and scalar field we take $A = A_t(r) dt$ and $\phi = \phi(r)$, respectively. Thus, $\xi \equiv \partial_t$ is a timelike Killing vector field and, since ${\cal L}_\xi \phi = {\cal L}_\xi A =0$, our solutions are globally static. Near the boundary of \AdS{4} ($r\to \infty$) we have the asymptotic expansion
\begin{equation}\label{eq:asyfalloffs}
\begin{gathered}
g(r) = r^2 + 1+ \frac{\phi_1^2}{2} -\frac{g_1}{r}+\ldots,\quad  \beta(r) = \beta_\infty + \ldots \\
A_t(r) = \mu - \frac{\rho}{r}+\ldots,\quad \phi(r) = \frac{\phi_1}{r} + \frac{\phi_2}{r^2}+\ldots 
\end{gathered}
\end{equation}
We will find it convenient to use the coordinate freedom in $t$ to set $\beta_\infty = 0$. 

We can read off various properties of the CFT from these asymptotics.  As mentioned, we can impose standard ($\phi_1$ fixed) or alternate ($\phi_2$ fixed) boundary conditions for the bulk scalar field, leading to the following identifications:
\begin{equation}
\begin{aligned}
\text{Standard (Dirichlet)}: & \quad \phi_2 = \vev{{\cal O}_2} \quad (\Delta_+=2) \\
\text{Alternate (Neumann)}:& \quad \phi_1 = \vev{{\cal O}_1} \quad (\Delta_-=1)
\end{aligned}
\end{equation}
The other quantities of interest in the boundary CFT are the conserved currents $J^\mu$ dual to the Maxwell field and the conserved boundary stress tensor $T^{\mu\nu}$.  The former is determined from the gauge field to be $\vev{J^\mu} = \rho \,\delta^\mu_{\;t}$ with $\mu$ the boundary source (chemical potential). The energy momentum tensor on the other hand receives contributions from the geometry as well as counter-terms involving the Maxwell and scalar fields \cite{Hartnoll:2008kx, Gentle:2011kv}.\footnote{ To be precise, imposing standard boundary conditions ($\phi_1 =0$) leads to no scalar contributions to the  boundary stress tensor, there being no counter-terms of interest. On the other hand, when $\phi_1 \neq 0 $ in alternate quantization we see that the metric functions get corrected due to the slow fall-off which necessitates scalar counter-terms as  originally described in \cite{Klebanov:1999tb}.} However, despite these differences the final result for the expectation value of the stress tensor is quite simple
\begin{equation}\label{eq:stresstensor}
\vev{T^\mu_{\phantom{\mu}\nu}} = \frac{g_1}{8\pi\, G_4}\; \text{diag}\left\{-1, \frac{1}{2}, \frac{1}{2}\right\}  = 2\, c_\text{eff}\, g_1\, \text{diag}\left\{-1, \frac{1}{2}, \frac{1}{2}\right\} 
\end{equation}
and is conserved and traceless (as it should be since there is no conformal anomaly in (2+1)-dimensional CFTs). In particular, the ADM mass of the solutions in either choice of scalar boundary condition is $M =g_1/(2\,G_4)$ and the minimal mass solution is simply AdS$_4$.

The  solitonic solutions we are interested in are basically ground states of the system with fixed scalar expectation value and charge. They can be constructed in a perturbative expansion around AdS$_4$.  For a given choice of quantization, we can choose the expectation value $\langle\mathcal{O}_{\Delta}\rangle \sim \varepsilon$ to be our small parameter.  The relevant metric functions are given in Appendix~\ref{app:pertubative}. Note that since the bulk scalar stress tensor is quadratic in the scalar field, the back-reaction on the metric occurs at $O(\varepsilon^2)$ in this perturbative expansion; this implies that in the perturbative limit $\vev{T^\mu_{\phantom{\mu}\nu}} \sim \varepsilon^2$. This simple observation that the response of the CFT degrees of freedom is non-linear in the scalar operator expectation value will play a role in our later discussion of relative entropy.

One can of course go beyond perturbation theory: fully back-reacted non-linear solutions of the system \eqref{eq:generalaction} can be found numerically.\footnote{ The simplest strategy is to integrate out the differential equations using a regular series expansion around the origin $r=0$,  integrate in using the asymptotic expansion \eqref{eq:asyfalloffs}, then match the two in the middle.}  Let us now review the results found in \cite{Gentle:2011kv}. For a given theory \eqref{eq:generalaction} the spectrum of solitons depends strongly on the details of the functions $\{V(\phi), Q(\phi)\}$. For  phenomenological models \eqref{eq:phenmodel} with fixed mass $m_\phi^2\,L^2 =-2$ there are
two classes of solution branch (independent of the scalar boundary conditions): \mbox{(i) a} branch that is connected to global \AdS{4}, part of which is accessible by perturbation theory and (ii) a branch that is entirely non-perturbative. The former class is characterised by bounded conserved charges for small $q$ and unbounded charges above a critical $q_c$, whereas the opposite is true for the latter class. See Fig.~7 in \cite{Gentle:2011kv} for results obtained by varying $q$.\footnote{ Qualitatively similar results in five dimensions were presented in \cite{Dias:2011tj}.}

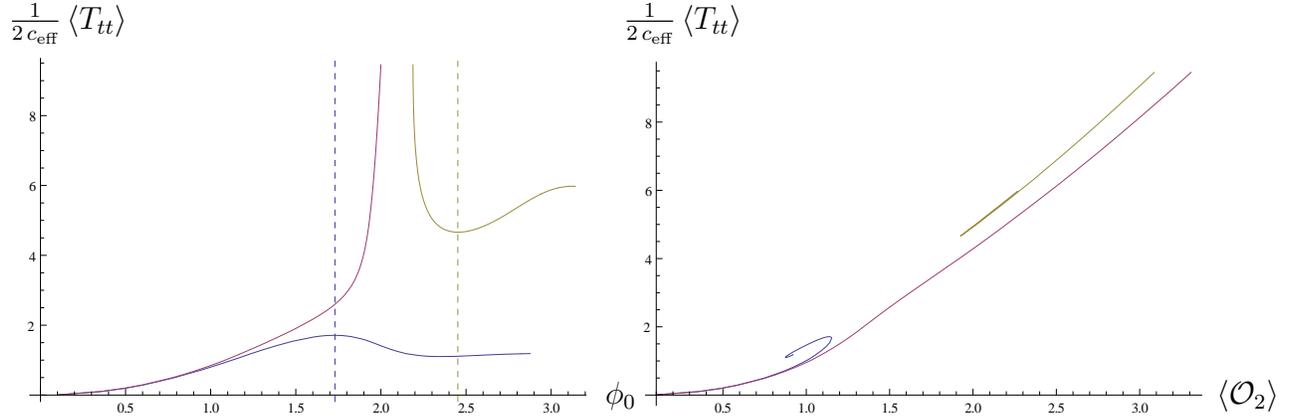
\begin{figure}[h!]
\vskip1.5em
\begin{center}
\includegraphics[width=0.45\textwidth]{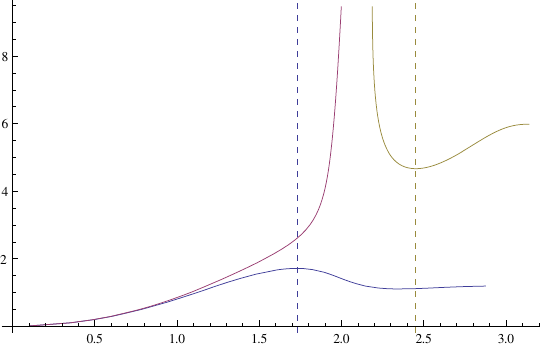}
\hskip1.5em
\includegraphics[width=0.45\textwidth]{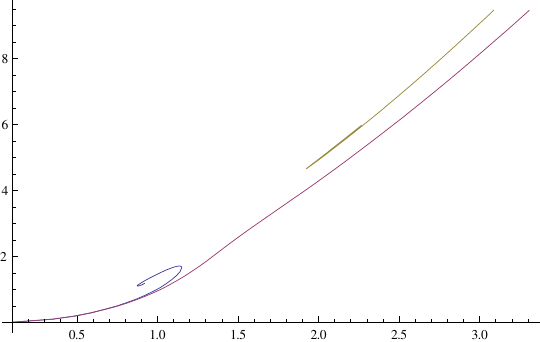}
\setlength{\unitlength}{0.1\columnwidth}
\begin{picture}(0.3,0.4)(0,0)
\put(-9.3,3.15){\makebox(0,0){$\frac{1}{2\, c_\text{eff}}\, \vev{T_{tt}}$}}
\put(-4.35,3.15){\makebox(0,0){$\frac{1}{2\, c_\text{eff}}\, \vev{T_{tt}}$}}
\put(-4.85,0.15){\makebox(0,0){$\phi_0$}}
\put(0.2,0.15){\makebox(0,0){$\vev{{\cal O}_2}$}}
\end{picture}
\end{center}
\vskip-0.5em
\caption{Branches of regular $\Delta=2$ solitons in the phenomenological model.  Here we plot the energy density against the core value $\phi_0$ of the bulk scalar (left) and the expectation value $\vev{{\cal O}_2}$ of the dual  operator (right). The dark blue and yellow branches both have $q^2=1.2$, whereas the magenta branch has $q^2=1.3$. The dark blue and yellow dashed lines denote the extrema $\phi_0^{\mathrm{max}}$ and $\phi_0^{\mathrm{min}}$, respectively, which will be useful reference points in \S\ref{sec:ee}.}\label{fig:HHHBranchesStandard}
\end{figure}

When the scalar charge is small $q<q_c$, in the bounded branch the ADM mass rises monotonically from zero as a function of the core value $\phi_0$ of the scalar field to a global maximum at some $\phi_0 = \phi_0^\text{max}$, then exhibits damped oscillations; see the blue curve of Fig.~\ref{fig:HHHBranchesStandard}.\footnote{ This oscillatory behavior is similar to that found for charged boson stars in flat space \cite{Jetzer:1989us}, neutral boson stars in AdS \cite{Astefanesei:2003qy} and radiation stars in \AdS{} \cite{Hubeny:2006yu}.  One expects that solutions before the first maximum will be stable to linearised homogeneous perturbations, whereas those with $\phi_0 > \phi_0^{\textrm{max}}$ will not, as is the case for the boson stars mentioned. While as far as we are aware this has not be checked explicitly for the solutions we are discussing here, it should be  possible to adapt the recent results of \cite{Green:2013ica,Schiffrin:2013zta} to confirm our suspicions.\label{fn:stability}} The unbounded branch for $q < q_c$ is characterized by being connected to the zero-temperature limit of charged hairy black holes in planar \AdS{}. For $q> q_c$ however we have a single physical branch of solutions that interpolates nicely from global AdS (the vacuum) to the ground state of planar holographic superconductors. In Fig.~\ref{fig:HHHBranchesStandard} we show examples of both types of connected branch as well as an unbounded disconnected branch for $q< q_c$.

In our second model, the $U(1)^4$ truncation, there are no free parameters.  A single bounded connected branch is found for the $\Delta=2$ quantization, whereas two types of unbounded branch are found for the $\Delta=1$ quantization: one connected and the other disconnected.\footnote{ The $\Delta =1 $ boundary condition is a supersymmetry preserving boundary condition in the theory. The connected branch of solutions in this case is the physical branch of solutions; the disconnected branch is sub-dominant in the micro-canonical ensemble with fixed energy and charge.} This theory also has a one-parameter family of singular analytical solitons that are neutral; these are curious in that their planar limit coincides with the planar limit of the  $\Delta=1$ connected unbounded branch. We discuss this family in Appendix~\ref{app:singular} as it displays some bizarre properties vis-\`{a}-vis entanglement and causal information.

\subsection{Holographic measures of information}\label{sec:holoinfo}%

Having described the  geometries we are dealing with and their dual CFT states, let us turn to the observables we would like to focus on. Let us consider a (2+1)-dimensional quantum field theory living on the Einstein static universe (ESU), ${\mathbb R} \times {\bf S}^2$.  This spacetime can be thought of as the conformal boundary $\partial\mathcal{M}$ of a (3+1)-dimensional static asymptotically AdS spacetime $\mathcal{M}$ with a metric of the form \eqref{eq:metricansatz}.  We are interested in the information content of states defined on a two-dimensional spatial region $\mathcal{A}\subset\partial\mathcal{M}$ that we choose to be the polar-cap $\mathcal{A}=\{(t,\theta,\varphi)\, |\, t = 0, |\theta|\leq\theta_{\mathcal{A}}\}$.  

One measure of information that we consider is the entanglement entropy  for $\mathcal{A}$. In the field theory this is the von Neumann entropy of the reduced density matrix $\rho_{\mathcal{A}}$ associated with $\mathcal{A}$. Following \cite{Ryu:2006bv}, it is computed holographically from the bulk theory via 
\begin{equation}\label{eq:holographicEE}
S_{\mathcal{A}} = \frac{\textrm{Area}\left[\extr{\cal A}\right]}{4\, G_4} \equiv 4\pi\,c_\text{eff}\,  \textrm{Area}\left[\extr{\cal A}\right]
\end{equation}
where $\extr{\cal A}$ is a bulk co-dimension two extremal surface in $\mathcal{M}$ anchored on $\entsurf$.  If this surface is not unique, we choose the one whose area is minimal among all such surfaces homologous to $\mathcal{A}$. Note that since we are discussing static states of the CFT, the extremal surface is in fact a minimal surface and lies on a constant $t$ slice.   

To find $\extr{\cal A}$ we parametrize them in terms of world-volume coordinates $\{s,\varphi\}$ and focus without loss of generality on the $t=0$ slice. The embedding of the surface is then given by $\theta=\theta(s)$, $r=r(s)$ and the area functional to minimize is simply
\begin{equation}\label{eq:areaequation}
\textrm{Area}\left[\extr{\cal A}\right]= 2\pi \int ds\, r\sin\theta \sqrt{\frac{1}{g(r)}\left(\frac{dr}{ds}\right)^2 + r^2\left(\frac{d\theta}{ds}\right)^2} \equiv 2\pi \int ds\, \mathcal{L}
\end{equation}
The Euler-Lagrange equations for $\theta(s)$ and $r(s)$ are equivalent due to reparametrization invariance of the area functional.  A further equation comes from the choice of the parameter $s$.  After imposing smoothness at $\theta=0$, we integrate these equations from the bulk point $ \rE\equiv r(\theta=0)$ and read off the $\theta_{\mathcal{A}}$ that this minimal surface is anchored on. 

Since the soliton spacetimes under consideration are causally trivial, we only need to consider $\theta_{\cal A} \in (0,\frac{\pi}{2})$ because minimal surfaces are symmetric under $\theta_{\cal A} \to \pi - \theta_{\cal A}$.  This in particular implies that for the region ${\cal A}$ being half of the boundary ${\bf S}^2$, i.e., $\theta_{\cal A} = \frac{\pi}{2}$,  the two-surface $\theta(r) = \frac{\pi}{2}$ that slices through the middle of the geometry is a minimal surface. We will henceforth denote this special surface as $\extr{\cal A}^\sharp$; it will prove useful in deriving some analytic expressions to orient our discussion.  Note that there may be other minimal surfaces with  $\theta_{\cal A}=\frac{\pi}{2}$ and it is not clear \emph{a priori} that this one has lowest area. In the cases where it does (which will transpire to be most of the physically relevant examples) one can immediately extract the entanglement entropy of one half of the boundary with the other. 

There is, however, more information contained in the density matrix.  It is Hermitian and positive so can be written in the form $\rho_{\mathcal{A}}\sim \exp\left(-H_{\cal A}\right)$, where  $H_{\cal A}$ is the modular or entanglement Hamiltonian.  Typically this operator is non-local and does not buy us much. However, in some special cases it is local and provides a novel measure of the energy contained in the causal development $\domd$ of the region ${\cal A}$.\footnote{ Note that since we are given the reduced density matrix $\rho_{\mathcal A}$ we can without loss of generality extend consideration to the boundary causal development or domain of dependence $\domd$ (which is a co-dimensional zero boundary region) associated with ${\cal A}$.}  For example, in any quantum field theory the modular Hamiltonian is proportional to a boost generator when $\domd$ is chosen to be the Rindler wedge of Minkowski space.

The local modular Hamiltonian for the Rindler wedge can be manipulated to provide a local modular Hamiltonian for the polar-cap regions of the ESU. To obtain this we  exploit the observation made in \cite{Casini:2011kv} that the causal development of the Rindler wedge can be conformally mapped to the casual development of the polar-cap of the ESU. It is in fact easy to show using the explicit map that the modular Hamiltonian takes the form 
\begin{equation}
H_{\cal A}=2\pi \int_{\mathcal{A}}d\Omega\; \frac{\cos\theta-\cos\theta_{\mathcal{A}}}{\sin\theta_{\mathcal{A}}}\,  T_{tt}(\Omega)
\end{equation}
where $T_{\mu\nu}$ is the stress tensor of the field theory.\footnote{ To derive this it suffices to note that the coordinate transformation 
$$t = \frac{\sin\theta_{\mathcal A} \, \sinh\tau}{\cosh u+ \cos\theta_{\cal A}\, \cosh \tau} \\, \quad \theta  = \frac{\sin\theta_{\mathcal A} \, \sinh u}{\cos\theta_{\cal A}\, \cosh u+ \cosh \tau}$$ 
confomally maps the causal development of the polar-cap region of ESU$_d$ into the Lorentzian hyperbolic cylinder  ${\mathbb R} \times H^{d-1}$ (which in turn is conformal to the Rindler wedge).
} 
This is a simple local operator and provides a measure of energy as measured by the reduced density matrix contained in the region of interest. For the soliton geometries described in \S\ref{sec:solitons} the boundary energy momentum tensor takes the form \eqref{eq:stresstensor} and we see that $\vev{T_{tt}}$ is  given by the (constant) ADM mass density. So for soliton states in the CFT we obtain 
\begin{equation}\label{eq:modHsoliton}
\langle H_{\cal A}\rangle= 4\pi^2 \left[ \int_0^{\theta_{\mathcal{A}}} d\theta\, \sin\theta\, \frac{\cos\theta-\cos\theta_{\mathcal{A}}}{\sin\theta_{\mathcal{A}}}\right] \frac{g_1}{8\pi\, G_4} = (4\pi)^2\, c_{\text{eff}}\, \frac{\sin^4\left(\theta_{\mathcal{A}}/2\right)}{\sin\theta_{\mathcal{A}}}\, g_1
\end{equation}
Therefore, up to a region-dependent factor the entanglement spectrum coincides with the mass spectrum of the solitons.

To appreciate the relevance of the modular Hamiltonian we recall the notion of the relative entropy. Given two density matrices $\rho_0$ and $\rho_1$ one defines 
\begin{equation}
S_\text{rel}\left(\rho_1;\rho_0\right)  \equiv - \rm{Tr} \left( \rho_1 \log \rho_1\right) + \rm{Tr} \left(\rho_1 \log  \rho_0 \right) 
\label{}
\end{equation}	
This relative entropy is manifestly positive-definite: $S_\text{rel} \geq 0$. For our considerations we will take $\rho_0$ to be the density matrix associated with the region ${\cal A}$ in the vacuum (global AdS geometry) and $\rho_1$ to be the one associated with the same region in a soliton geometry. 

Motivated by this concept, let us define two quantities that will be of interest in what follows. We first define the entanglement contained in the state relative to the vacuum (focusing on a particular CFT region ${\cal A}$):
\begin{equation}
\Delta S_{\cal A}  \equiv
S_{\cal A}^{\vev{{\cal O}_\Delta}} - S_{\cal A}^{\text{vacuum}}
\label{}
\end{equation}	
By construction, $\Delta S_{\cal A}$ is UV-finite since the leading divergences are independent of the state of the system. We can similarly define the modular Hamiltonian expectation value relative to the vacuum via
\begin{equation}
\Delta H_{\cal A}\equiv \vev{H_{\cal A}^{\vev{{\cal O}_\Delta}}} - \vev{H_{\cal A}^\text{vacuum}} = \vev{H_{\cal A}^{\vev{{\cal O}_\Delta}}}
\label{}
\end{equation}	
where we have applied the result \eqref{eq:modHsoliton} for $m=0$ AdS$_4$. Since the theory \eqref{eq:generalaction} satisfies the positive energy theorem in AdS, we have $\Delta H_{\cal A} \geq 0$.  Furthermore, using the positivity of relative entropy introduced above, as argued in \cite{Blanco:2013joa} we are guaranteed that 
\begin{equation}
S_\text{rel}\left(\rho_{\cal A}^{\vev{{\cal O}_\Delta}}; \rho_{\cal A}^\text{vacuum}\right) \geq 0 \qquad \Longrightarrow \qquad \Delta H_{\cal A} \geq \Delta S_{\cal A}
\label{eq:positiverelativeentropy}
\end{equation}	
In addition, the inequality is saturated at leading order in the deviation from the vacuum. For the solitonic coherent states, however, since the deformation in the geometry is at least quadratic in the vacuum expectation value of the operator, we therefore have
\begin{equation}
\Delta H_{\cal A} = \Delta S_{\cal A}  =0 \quad \text{to linear order in} \;\;\vev{{\cal O}_\Delta}
\label{}
\end{equation}	
So the observables we would be focusing on will start capturing the effects of the coherent state only at the non-linear order.

Another measure of information is derived from the bulk causal wedge $\cwedge$ associated with ${\cal A}$. It is defined to be  set of bulk points that can influence and be influenced by points on the boundary domain of dependence $\domd$.  While the causal wedge is itself a bulk co-dimension zero volume, its boundary $\bcwedge$ in the bulk is generated by null geodesics that end on $\domd$. 
One can intuitively view this bulk null surface $\bcwedge$ as the union of two null surfaces that correspond to the past and future directed geodesics, i.e., $\bcwedge = \bcwedgep \cup \bcwedgef$.  At the intersection   there lies a bulk co-dimension two surface called the  causal information surface $\csf{\cal A}$ whose area in Planck units is the causal holographic information $\chi_{\cal A}$  \cite{Hubeny:2012wa}:
\begin{equation}
\chi_{\cal A}=  \frac{\textrm{Area}\left[\csf{\cal A}\right]}{4\, G_4} \equiv 4\pi\,c_\text{eff}\,  \textrm{Area}\left[\csf{\cal A}\right]
\label{}
\end{equation}	
Note that $\csf{\cal A}$ is the minimal area such surface on $\bcwedge$.
While there is no clear understanding of $\chi_{\cal A}$ from field theory yet  (see however the interesting recent proposal of \cite{Kelly:2013aja} and earlier attempts by \cite{Freivogel:2013zta}), as emphasized in the original construction the naturalness of causal constructions makes it an interesting quantity to consider from the bulk perspective.

It is again useful to monitor the relative causal holographic information in a given region by subtracting off the vacuum answer. It has been shown in 
\cite{Hubeny:2012wa,  Wall:2012uf, Hubeny:2013gba} that the causal holographic information generically differs from the entanglement entropy, with $\chi_{\cal A} > S_{\cal A}$. More importantly, the UV divergence structure of  $\chi_{\cal A}$ is stronger for an arbitrarily-shaped region (even in the vacuum state). However,  it was also demonstrated in \cite{Hubeny:2012wa} that for polar-cap regions of the vacuum state of the CFT in global \AdS{d+1} the two concepts coincide: $\chi_{\cal A}^\text{vacuum} =S_{\cal A}^\text{vacuum}$. This implies that 
 \begin{equation}
\Delta \chi_{\cal A} \equiv
\chi_{\cal A}^{\vev{{\cal O}_\Delta}} - \chi_{\cal A}^{\text{vacuum}} = 
\chi_{\cal A}^{\vev{{\cal O}_\Delta}} - S_{\cal A}^{\text{vacuum}}
 \label{delchiA}
 \end{equation}	
We further anticipate that $\Delta \chi_{\cal A}\geq \Delta S_{\cal A}$ in light of the above.

As mentioned earlier there is an interesting interplay between the two surfaces $\csf{\cal A}$ and $\extr{\cal A}$ in the bulk geometry. What is however more curious is that the causal wedge itself can have non-trivial topology despite ${\cal A}$ being simply connected. Clearly non-trivial topology for the causal wedge translates into the fact that the causal information surface $\csf{\cal A}$ comprises of disconnected examples.  This was illustrated explicitly in \cite{Hubeny:2013gba} for the global Schwarzschild-\AdS{5} black hole. 

While this explicit demonstration was in a causally non-trivial spacetime (indeed one of the components of the surface $\csf{\cal A}$ straddles the bifurcation surface of the black hole), it was argued there that the phenomenon is more generic and should persist even in the absence of bulk horizons. The essential point is that the non-trivial topology is a consequence of the steep gravitational potential in the bulk and one anticipates that this can be achieved even in the absence of a black hole. It was furthermore conjectured that the 
in spherically symmetric spacetimes, an essential requirement for the causal wedge to develop holes is that the geometry must admit null circular orbits. In what follows we will investigate this phenomenon in the causally trivial soliton spacetimes and show that 
they do entertain this phenomenon explicitly. In fact, consistent with the conjecture of \cite{Hubeny:2013gba} we will find that  non-trivial topology of the  causal wedge is precisely correlated with the presence of null circular orbits in the spacetime.

\section{Minimal surfaces and entanglement entropy}
\label{sec:ee}%

We begin our discussion with a study of minimal surfaces anchored on the polar-cap in   regular soliton geometries.\footnote{ Results for the singular soliton found in the $U(1)^4$ truncation are described in Appendix~\ref{app:singular} for completeness. It is worthwhile remarking here that the singular soliton has a time-like singularity at its core and so shares features with the unphysical negative mass Schwarzschild-AdS solution. In particular, the singularity {\em attracts} the extremal surfaces towards it, implying that the  entanglement relative to the vacuum is non-positive definite $\Delta S_{\cal A} \leq 0$. Also the causal properties of the solution are bizarre: we see a gravitational {\em time-advance} phenomenon --- it is faster to communicate between boundary points through the bulk! We return to the latter issue in \S\ref{sec:wedges}.}
To orient ourselves, let us warm up by studying minimal surfaces in global AdS$_4$, for which $g(r)=r^2+1$ and $\beta(r)=0$ in \eqref{eq:metricansatz}. As is well known, such surfaces can be found analytically; one has
\begin{equation}\label{eq:AdSthetaofr}
\theta(r) = \cot^{-1} \sqrt{\frac{r^2+1}{\left(r/\rE^{\mathrm{AdS}}\right)^2-1}} \to
\left\{ 
\begin{array}{ c l }
0 & \textrm{as}\ r\to \rE^{\mathrm{AdS}}\\
\theta_{\mathcal{A}}=\cot^{-1}\rE^{\mathrm{AdS}} &\textrm{as}\ r\to \infty
\end{array}
\right.
\end{equation}
This family of surfaces foliates the entire geometry; in other words, one can construct a minimal surface extending to any desired $\rE^{\mathrm{AdS}}\geq 0$ if a suitable choice of $\theta_{\mathcal{A}}$ is made. We will show later that it is not always possible to foliate the constant time slices of a given geometry of the form \eqref{eq:metricansatz} with smooth minimal surfaces even in causally trivial geometries.\footnote{ For causally non-trivial spacetimes, such as global Schwarzschild-AdS, the results of \cite{Hubeny:2013gta} demonstrate the absence of such a foliation.}

Let us now turn to the soliton geometries. We now have non-vanishing bulk matter fields leading to a deeper gravitational potential near the core (relative to the vacuum). Wandering into the gravity well extracts an area price, as is clear from \eqref{eq:areaequation}. All other things being equal, this would lead us to expect that the soliton core will {\em repel}
the minimal surfaces $\extr{\cal A}$ and so they will not reach as far into the bulk as in AdS$_4$ for given $\theta_{\mathcal{A}}$ in these geometries, i.e., $\rE>\rE^{\mathrm{AdS}}$.  We will see that this expectation is borne out if the scalar field obeys the Dirichlet boundary condition (standard quantization) at infinity, but not if it obeys the Neumann boundary condition (alternate quantization). In the latter case we shall see that the reason has to do with modifications to the {\em asymptotic} gravitational potential.\footnote{ A word of caution: we use the phrase gravitational potential to encode the information contained in $g_{rr}$ and refer to the temporal component of the metric $g_{tt}$ as the red-shift factor (eschewing the neologism {\em emblackening} for obvious reasons).} 

Without further ado let us now turn to the behaviour of the entanglement entropy in the two distinct theories.

\subsection{Dirichlet boundary conditions $\vev{{\cal O}_2} \neq 0$}
\label{sec:standard}

\begin{figure}[!htb]
\begin{center}
\includegraphics[width=0.35\textwidth]{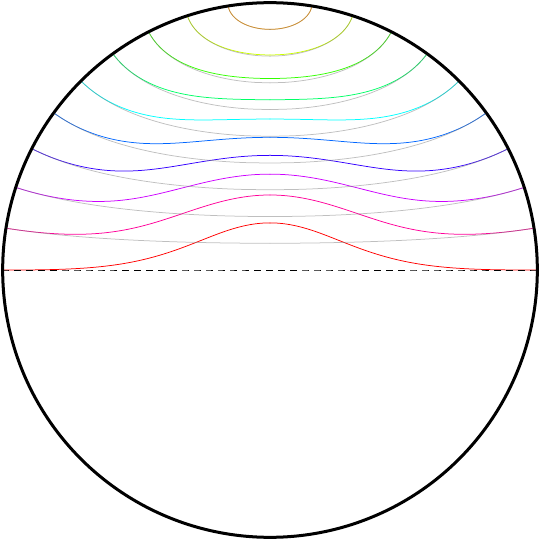}
\hskip1em
\includegraphics[width=0.5\textwidth]{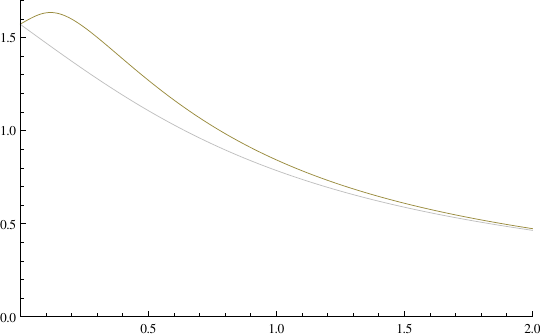}
\setlength{\unitlength}{0.1\columnwidth}
\begin{picture}(0.3,0.4)(0,0)
\put(-4.95,3.4){\makebox(0,0){$\theta_{\mathcal{A}}$}}
\put(0.15,0.15){\makebox(0,0){$\rE$}}
\end{picture}
\end{center}
\vskip-0.5em
\caption{Left: Minimal surfaces in a $\Delta=2$  soliton  on the unbounded disconnected branch with $\phi_2=1.93(9)$ in the phenomenological model. Curves correspond to $\theta_{\mathcal{A}}=\frac{\pi}{20},\frac{2\pi}{20},\ldots,\frac{\pi}{2}$ from top to bottom, with surfaces anchored at the same points in AdS$_4$ shown in grey and the symmetric surface $\extr{\cal A}^\sharp$ shown as a dashed black line. Right: Minimal surfaces in the same geometry, with the AdS$_4$ curve $\theta_{\mathcal{A}}=\cot^{-1}\rE^{\mathrm{AdS}}$ shown in grey. Note that surfaces with $\theta_{\cal A}>\pi/2$ can be mapped to $\theta_{\cal A}<\pi/2$ using the aforementioned symmetry.}
\label{fig:HHHType3}
\end{figure}
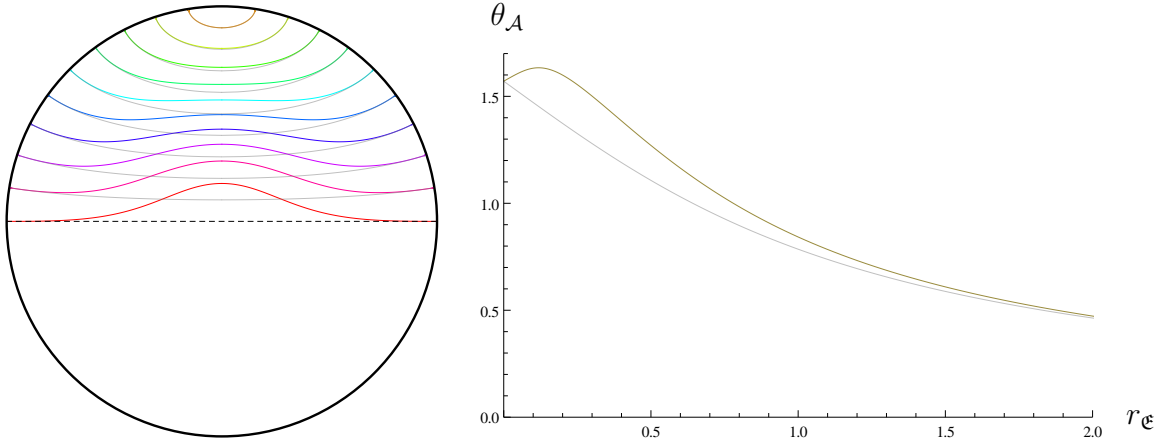

To describe the behaviour of the minimal surfaces when we demand that the scalar field $\phi$ satisfies $\phi_1 =0$ at infinity, we consider for definiteness\footnote{ We expect that the behaviour in the $U(1)^4$ truncation with these Dirichlet boundary conditions will be qualitatively similar.} the phenomenological model \eqref{eq:phenmodel}. 
For all branches shown in Fig.~\ref{fig:HHHBranchesStandard}, we find the anticipated repulsion of minimal surfaces from the core.  The detailed behaviour of course depends on the particular branch under consideration and the core value  $\phi_0$. More specifically:

\begin{itemize}
\item The minimal surfaces in unbounded connected solitons (magenta branch of Fig.~\ref{fig:HHHBranchesStandard}) are very similar to those in AdS$_4$, as are those for bounded connected solitons (dark blue branch of Fig.~\ref{fig:HHHBranchesStandard}) with $\phi_0<\phi_0^{\mathrm{max}}$.  Curves in the $(\rE,\theta_{\mathcal{A}})$ plane decrease monotonically from $(0,\frac{\pi}{2})$ and lie above the AdS$_4$ curve. 
\item However, things look rather different for solitons further along the connected bounded branch and all along the unbounded disconnected branch (yellow branch of Fig.~\ref{fig:HHHBranchesStandard}) --- see Fig.~\ref{fig:HHHType3} for minimal surfaces found in a soliton on the latter branch. The repulsion effect is quite pronounced and now we find multiple minimal surfaces for a given $\theta_{\mathcal{A}}$.  The value $\phi_0^{\mathrm{h}}$ above which this multiplicity occurs is strictly greater than $\phi_0^{\mathrm{max}}$ for the connected branch and strictly less than $\phi_0^{\mathrm{min}}$ for the disconnected branch.\footnote{ In fact, it appears that $m(\phi_0^{\mathrm{h}})$ equals the value of $m$ at the first maximum in the latter curve.}
\end{itemize}

To get a feeling for the entanglement relative to the vacuum, it is useful to examine what happens for small values of $\vev{{\cal O}_2} = \varepsilon$. The first correction to the entanglement entropy is given by evaluating the area of the vacuum minimal surface,  \eqref{eq:AdSthetaofr}, in the perturbative soliton geometry described in Appendix~\ref{app:pertubative}. We can compute this area analytically for general~${\cal A}$ in the particular case of $\Delta=2$, giving the following regulated entanglement entropy:
\begin{align}\label{eq:pertareadelta2}
\frac{1}{4\pi\,c_\text{eff}}\, \Delta {S}_{\cal A}  &=   \int_{\cot\theta_{\cal A}}^{\infty} dr\, \frac{\pi \left(r^2-\cot^2\theta_{\cal A}\right) \left[\left(r^2+1\right) \tan ^{-1}r-r\right]\sin\theta_{\cal A}}{2r^2 \left(r^2+1\right)^{5/2}}\, \varepsilon^2 + O(\varepsilon^4) \nonumber\\
&= \frac{\pi}{24\, \sin\theta_{\cal A}} \left[9 \pi+ 
   12 \left( 2 \theta_{\cal A}-\pi \right)\cos\theta_{\cal A} + 
   3 \pi \cos 2 \theta_{\cal A} - 18 \sin \theta_{\cal A}- 2 \sin 3 \theta_{\cal A}\right]\varepsilon^2\nonumber\\ 
   &\phantom{=\ }+ O(\varepsilon^4)
   \end{align}
The fact of import for the moment is that the coefficient of $\varepsilon^2$ is positive definite, taking its maximum value when $\theta_{\cal A}=\frac{\pi}{2}$. Using \eqref{eq:modHsoliton} for the modular Hamiltonian and the pertubative result  $g_1=\frac{\pi}{4}\,\varepsilon^2+ O(\varepsilon^4)$  from \cite{Gentle:2011kv}, we find that $\Delta H_{\cal A} > \Delta S_{\cal A}$ for all $\theta_{\cal A}\in (0,\frac{\pi}{2}]$, as expected from the general argument based on the relative entropy \cite{Blanco:2013joa}.

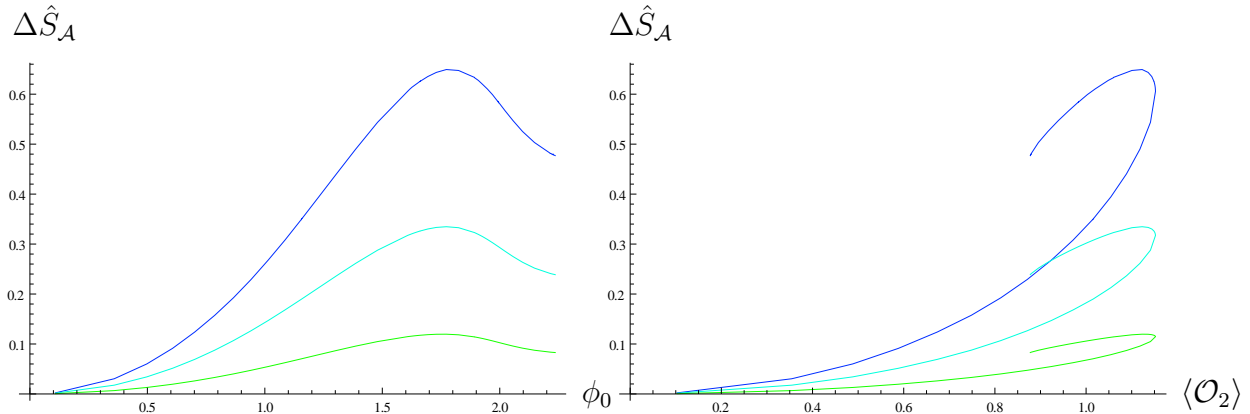
\begin{figure}[h!]
\vskip1.5em
\begin{center}
\includegraphics[width=0.45\textwidth]{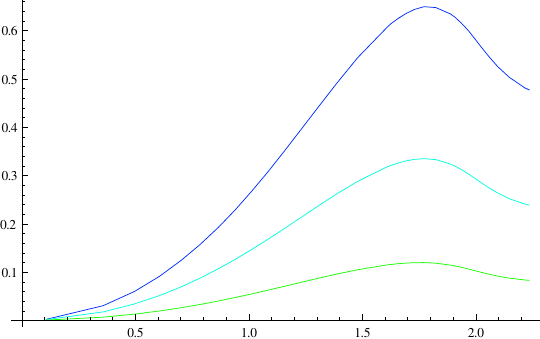}
\hskip1em
\includegraphics[width=0.45\textwidth]{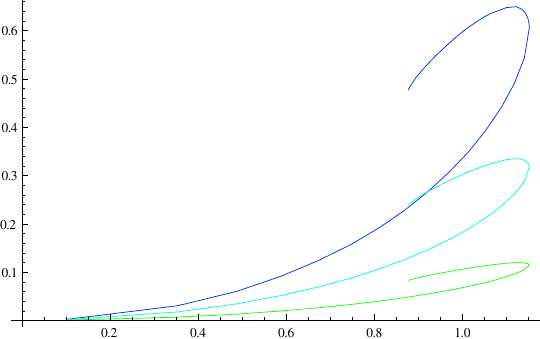}
\setlength{\unitlength}{0.1\columnwidth}
\begin{picture}(0.3,0.4)(0,0)
\put(-9.2,3.15){\makebox(0,0){$\Delta \hat{S}_{\cal A}$}}
\put(-4.4,3.15){\makebox(0,0){$\Delta \hat{S}_{\cal A}$}}
\put(-4.75,0.15){\makebox(0,0){$\phi_0$}}
\put(0.2,0.15){\makebox(0,0){$\vev{{\cal O}_2}$}}
\end{picture}
\end{center}
\caption{Entanglement relative to the vacuum (rescaled) in $\Delta=2$ solitons on the bounded connected branch in the phenomenological model. From top to bottom: $\theta_{\mathcal{A}}=1,0.75,0.5$.}
\label{fig:HHHdim2Areas}
\end{figure}

The general behaviour as a function of $\vev{{\cal O}_2}$ is straightforward to obtain using the numerical solutions. The only technical issue is the evaluation of the regulated areas, since the soliton geometries are known numerically up to some radial cut-off $r=R$.  
To obtain $\Delta S_{\cal A}$ for surfaces with non-trivial embedding profiles, we fix 
$\theta(r=R)=\theta^R_{\mathcal{A}}$. Computing the area of the minimal surface in the soliton geometry up to this chosen cut-off as a function of $\theta_{\cal A}$, we 
subtract off the area of the corresponding surface with the same cut-off  in \AdS{4}. Using the explicit profile \eqref{eq:AdSthetaofr} it is easy to see that the latter is in fact just $\textrm{Area} \left[\extr{\cal A}(R)\right] = 2\pi  \left(-1+\sqrt{1+R^2\sin^2\theta^R_{\mathcal{A}}}\right)$. For ensuring numerical convergence it turns out to be effective to use a parameterization of the surface such that 
 $\mathcal{L}_{\mathrm{on-shell}}= r\sin\theta$.\footnote{ We thank Henry Maxfield for this suggestion.} Taking this into account we can compute $\Delta S_{\cal A}$ as a function of $ \phi_0$ or $\vev{{\cal O}_2}$; in Fig.~\ref{fig:HHHdim2Areas} we show examples for $\Delta=2$ solitons in the phenomenological model. The regulated area follows the qualitative shape of the mass curves shown in Fig.~\ref{fig:HHHBranchesStandard} for all branches.\footnote{ To avoid notational clutter in the figures we henceforth denote physical quantities rescaled by the central charge with a hat: $\{
 \hat{S}_{\cal A}, \hat{H}_{\cal A},\hat{\chi}_{\cal A}\}  \equiv\frac{1}{4\pi\,c_\text{eff}}\,  \{S_{\cal A}, H_{\cal A},\chi_{\cal A}\}$.} 
 The behaviour as a function of the region size $\theta_{\cal A}$ is simple:
increasing $\theta_{\cal A}$ causes $\rE$ to decrease and thus $\Delta S_{\cal A}$ to increase.  

In all examples examined we confirmed the expectation that $\Delta H_{\cal A} > \Delta S_{\cal A}$ for generic $\vev{{\cal O}_2}$ as guaranteed by the positivity of relative entropy \eqref{eq:positiverelativeentropy}. The positive energy theorem for \eqref{eq:generalaction} with $\phi_1 =0$ ensures that $\Delta H_{\cal A} > 0$. Likewise the entanglement in these coherent states relative to the vacuum is also positive definite. In most regards these states of the CFT behave as one naively expects: creating the fine-tuned state requires a tighter entanglement of the degrees of freedom and so extracts an energy cost.

Before we move on, we should note that minimal surfaces in  soliton geometries similar to the bounded connected type arising in the phenomenological model were also studied earlier in \cite{Nogueira:2013if}. As noted above, in this case we have multiple minimal surfaces for a given $\theta_{\cal A}$  when the core scalar is large enough, i.e., when  $\phi_0>\phi_0^{\mathrm{h}}$. In all these cases consistent with their analysis we find that the surface with smallest $\rE$, i.e., the one that reaches the furthest into the bulk, has greater area relative to the one that sits closer to the boundary. This makes sense from our intuition based on the gravitational potential at the core. However, it is the dominant saddle point of the area functional that contributes to the entanglement entropy.
Consequently, in these spacetimes, the families of minimal surfaces that are the true minima of the area functional do not foliate constant time slices, leaving a `hollow' region around the core.\footnote{ A necessary corollary of this statement is that the simple analytical minimal surface $\extr{\cal A}^\sharp$ ceases to be the dominant saddle in these examples for $\theta_{\cal A} = \frac{\pi}{2}$.}

We observe a similar phenomenon in the unbounded disconnected soliton branch as well. In both these cases the region in soliton solution space where the phenomenon is manifest is one where the bulk solutions are likely to be unstable, as mentioned in footnote \ref{fn:stability}. Indeed we anticipate that these geometries do not actually correspond to stable CFT coherent states; for fixed $\vev{{\cal O}_2}$ the bulk geometry of relevance should be free of such exotic behaviour (see foonote \ref{fn:stability}).  

\subsection{Neumann boundary conditions $\vev{{\cal O}_1} \neq 0$}
\label{sec:alternate}

\begin{figure}[!htb]
\begin{center}
\includegraphics[width=0.35\textwidth]{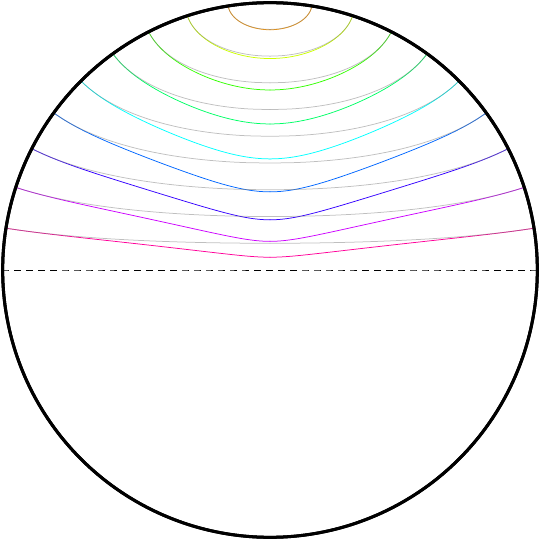}
\hskip2em
\includegraphics[width=0.5\textwidth]{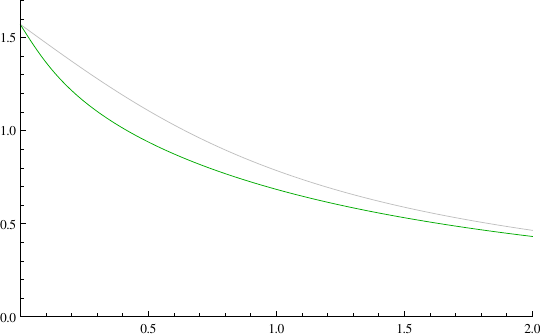}
\setlength{\unitlength}{0.1\columnwidth}
\begin{picture}(0.3,0.4)(0,0)
\put(-4.95,3.4){\makebox(0,0){$\theta_{\mathcal{A}}$}}
\put(0.15,0.15){\makebox(0,0){$\rE$}}
\end{picture}
\end{center}
\caption{Left: Minimal surfaces in a $\Delta=1$ soliton on the unbounded connected branch with $\phi_1=2.00(9)$ in the $U(1)^4$ truncation.  Curves correspond to $\theta_{\mathcal{A}}=\frac{\pi}{20},\frac{2\pi}{20},\ldots,\frac{9\pi}{20}$ from top to bottom, with surfaces anchored at the same points in AdS$_4$ shown in grey and the symmetric surface $\extr{\cal A}^\sharp$ shown as a dashed black line. Right: Minimal surfaces in the same geometry, with the AdS$_4$ curve $\theta_{\mathcal{A}}=\cot^{-1}\rE^{\mathrm{AdS}}$  shown in grey.}
\label{fig:DGSurfaces}
\end{figure}
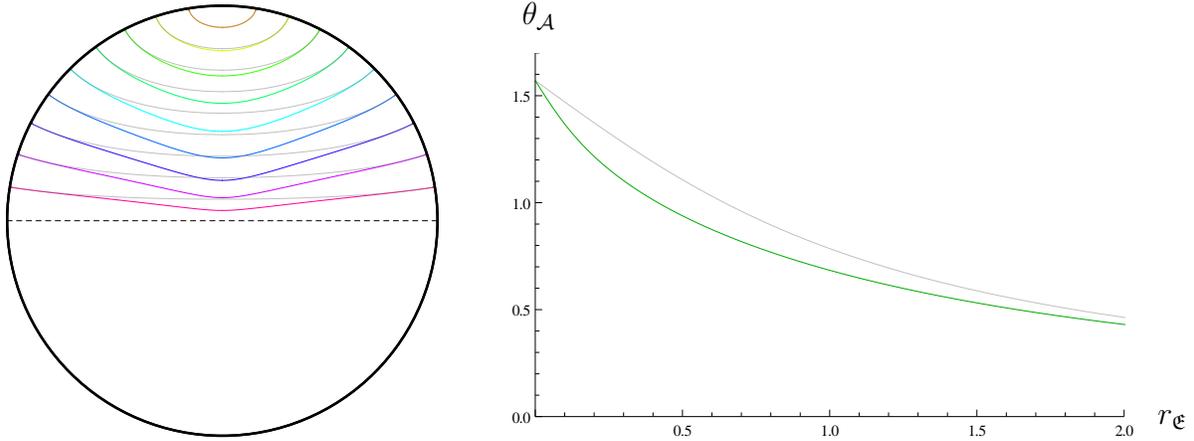

While our analysis of coherent states in the Dirichlet boundary condition case bore out most of our naive expectations, the Neumann boundary condition offers surprises. Consider then first the minimal surfaces in the solitons where $\phi_2 = 0$. In Fig.~\ref{fig:DGSurfaces} we present minimal surfaces and $(\rE,\theta_{\mathcal{A}})$ curves  in regular $\Delta=1$ solitons in the $U(1)^4$ truncation. We clearly see that the minimal surfaces in these geometries reach further into the bulk than in AdS$_4$ for given $\theta_{\mathcal{A}}$, i.e., $\rE<\rE^{\mathrm{AdS}}$. They appear to be `attracted' to the core of the soliton, despite the geometry being regular there.\footnote{ We have chosen to switch models because the behaviour of the minimal surfaces is more striking.  We return to the phenomenological model at the end of this section.}

This counter-intuitive behaviour can be understood as follows. Recall our argument that minimal surfaces prefer to stay away from steep gravitational potentials. Introducing matter into AdS in a spherically symmetric fashion, we usually expect an increased potential at the core of the soliton. However, when we relax the scalar boundary condition to allow for $\phi_1 \neq 0$ (and demand $\phi_2 =0$ as here) the asymptotic form of the metric changes as well --- see \eqref{eq:asyfalloffs}. The operative point is that the metric function $g(r)$ is deformed at ${\cal O}(r^0)$ instead of ${\cal O}(r^{-1})$, so in effect there is a greater potential at the asymptotic end than what would be encountered in the vacuum AdS spacetime.\footnote{ Despite the change in the metric at an earlier order in the large $r$ expansion, the solutions are nevertheless asymptotically globally AdS.} This increased potential at the boundary end causes the surfaces to migrate away from there to minimize their area. In general the increased potential at infinity is the dominant effect; even for large scalar core values, the surfaces keep being attracted to the core of the soliton.

As the minimal surfaces wander deeper into the bulk, their areas must correspondingly decrease relative to that of \AdS{4}. To see this explicitly, consider first  the perturbative solitons with $\vev{{\cal O}_1} = \varepsilon$. We can again evaluate the area integral and find that the regulated entanglement entropy is
\begin{align}\label{eq:pertarea}
\frac{1}{4\pi\, c_\text{eff}}\,\Delta S_{\cal A}&=  \int_{\cot\theta_{\cal A}}^{\infty} dr\, \frac{\pi \left(r^2-\cot^2\theta_{\cal A}\right)  \left(\tan ^{-1}r-r\right)\sin\theta_{\cal A}}{2r^2 \left(r^2+1\right)^{3/2}}\, \varepsilon^2 + O(\varepsilon^4) \nonumber\\
&=  \frac{\pi}{8\, \sin\theta_{\cal A}} \left[3 \pi +4 \left( 2 \theta_{\cal A}-\pi \right)\cos\theta_{\cal A} + 
    \pi \cos 2 \theta_{\cal A} - 8 \sin \theta_{\cal A}\right]\varepsilon^2+ O(\varepsilon^4)
\end{align}
Here, the coefficient of $\varepsilon^2$ is \emph{negative}, as we suspected.  It takes its minimum value when $\theta_{\cal A}=\frac{\pi}{2}$.

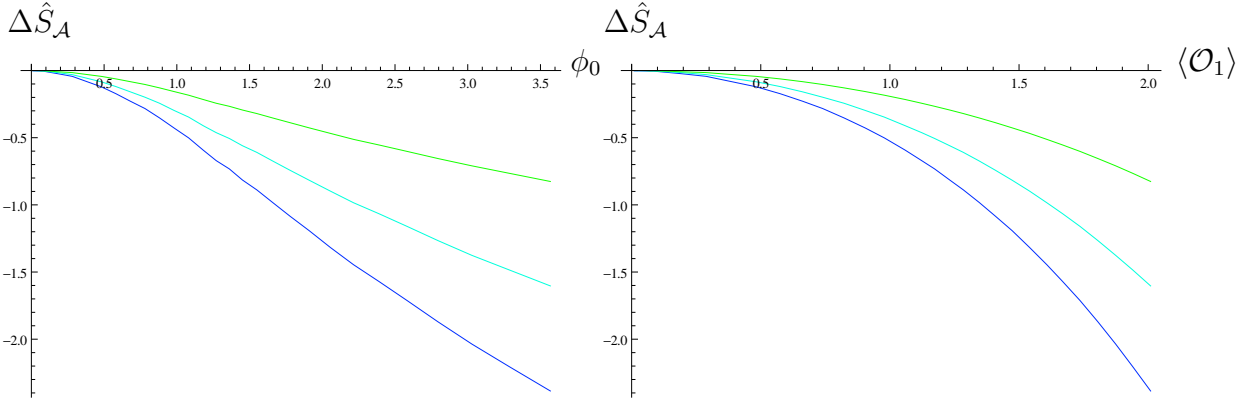
\begin{figure}[h!]
\vskip2em
\begin{center}
\includegraphics[width=0.45\textwidth]{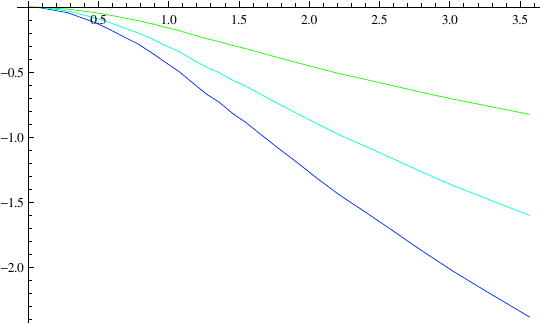}
\hskip1em
\includegraphics[width=0.45\textwidth]{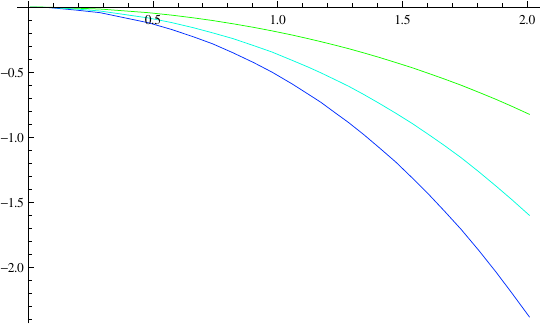}
\setlength{\unitlength}{0.1\columnwidth}
\begin{picture}(0.3,0.4)(0,0)
\put(-9.2,3.05){\makebox(0,0){$\Delta \hat{S}_{\cal A}$}}
\put(-4.4,3.05){\makebox(0,0){$\Delta \hat{S}_{\cal A}$}}
\put(-4.8,2.7){\makebox(0,0){$\phi_0$}}
\put(0.2,2.7){\makebox(0,0){$\vev{{\cal O}_1}$}}
\end{picture}
\end{center}
\caption{Entanglement relative to the vacuum (rescaled) in $\Delta=1$ solitons on the unbounded connected branch in the $U(1)^4$ truncation. From bottom to top: $\theta_{\mathcal{A}}=1,0.75,0.5$.}
\label{fig:DGarea}
\end{figure}

At first sight this seems very strange: this means that the atypical pure state dual to this soliton has \emph{lower} entanglement entropy than the vacuum.  However, this result persists in all examples of $\Delta=1$ solitons we have studied in both theories, for  general $\extr{\cal A}$; see Fig.~\ref{fig:DGarea} for examples in the $U(1)^4$ truncation.
Of course, the modular Hamiltonian for these solitons is positive definite and the relation $\Delta H_{\cal A} > \Delta S_{\cal A}$ is trivially satisfied. We find it curious that the coherent states of ${\cal O}_1$ add energy to the vacuum whilst lowering the entanglement. 

At this point the reader may rightly be skeptical about our subtraction scheme. Clearly, by adding a boundary term in the entanglement entropy given by a functional of $\vev{{\cal O}_1}$ one can raise $\Delta S_{\cal A}$ to a positive value. Given that such boundary terms are present in the boundary stress tensor, perhaps we are missing these contributions in the entanglement entropy. This is in fact not the case: it is easy to show that there cannot be additional boundary contributions to the area integral using the generalized entropy argument of \cite{Lewkowycz:2013nqa}. 
We will return to this point and some physical consequences of this phenomenon (and its generalizations) in \S\ref{sec:discussion}.

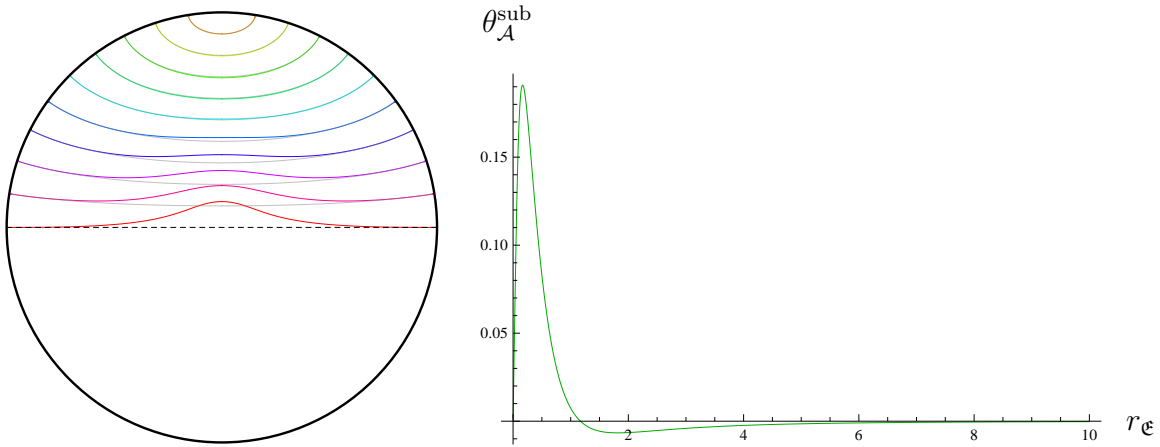
\begin{figure}[!htb]
\vskip1em
\begin{center}
\includegraphics[width=0.35\textwidth]{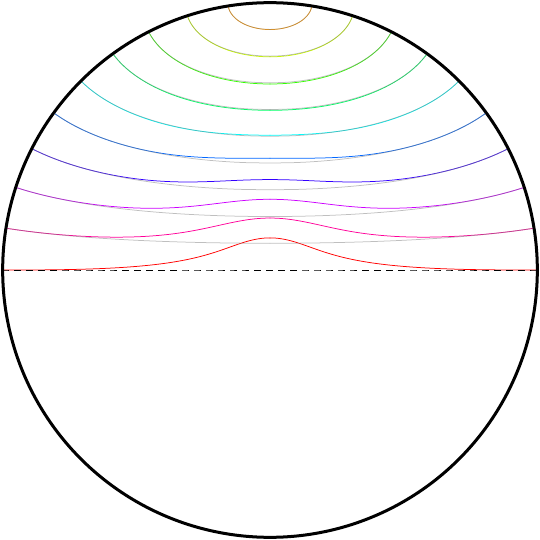}
\hskip1em
\includegraphics[width=0.5\textwidth]{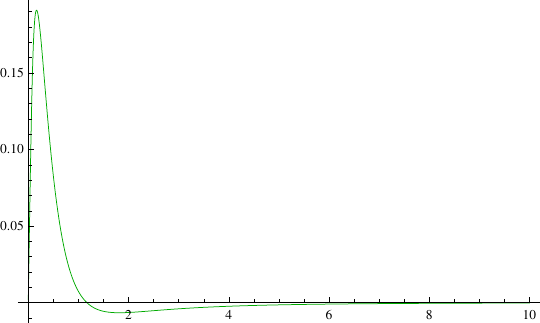}
\setlength{\unitlength}{0.1\columnwidth}
\begin{picture}(0.3,0.4)(0,0)
\put(-4.95,3.4){\makebox(0,0){$\theta_{\mathcal{A}}^{\mathrm{sub}}$}}
\put(0.15,0.15){\makebox(0,0){$\rE$}}
\end{picture}
\end{center}
\vskip-0.5em
\caption{Left: Minimal surfaces in a $\Delta=1$  soliton  on the bounded connected branch with $\phi_1=1.57(4)$ in the phenomenological model. Curves correspond to $\theta_{\mathcal{A}}=\frac{\pi}{20},\frac{2\pi}{20},\ldots,\frac{\pi}{2}$ from top to bottom, with surfaces anchored at the same points in AdS$_4$ shown in grey and the symmetric surface $\extr{\cal A}^\sharp$ shown as a dashed black line. Right: Minimal surfaces in the same geometry, with the AdS$_4$ curve $\theta_{\mathcal{A}}=\cot^{-1}\rE^{\mathrm{AdS}}$ subtracted.}
\label{fig:HHHType1}
\end{figure}

We note in passing that in contrast to the $\Delta=2$ solitons in the previous subsection, here both $\Delta S_{\cal A}$  and $\rE$ decrease with region size.  The spatial sections of these geometries can thus be foliated by smooth minimal surfaces. Also, see Fig.~\ref{fig:HHHType1} for minimal surfaces found in a $\vev{{\cal O}_1}$ soliton on the bounded connected  branch in the phenomenological model. Surfaces anchored at large $\theta_{\mathcal{A}}$ are repelled from the core, just as for $\vev{{\cal O}_2}$ solitons.  However, the opposite is true for small $\theta_{\mathcal{A}}$, just as for $\vev{{\cal O}_1}$ solitons in the $U(1)^4$ truncation.  (This feature is clearer in right-hand plot, but is also visible in the left-hand plot at high zoom.) This shows that there can be a subtle competition between the core and asymptotic potentials, but we emphasize that the result of entanglement reduction for $\Delta=1$ is unchanged.\footnote{ See Fig.~\ref{fig:chidelta1} for $\Delta S_{\cal A}$  in $\vev{{\cal O}_1}$ solitons in the phenomenological model.}

\section{Causal wedges in the soliton geometries}
\label{sec:wedges}%

We now turn to a slightly different measure of holographic information and look at causal wedges in our soliton geometries.  To construct the bulk causal wedge we first need to ascertain the boundary domain of dependence. For the polar-cap regions of interest, i.e., $\mathcal{A}=\{(t,\theta,\varphi)\, |\, t = 0, |\theta|\leq\theta_{\mathcal{A}}\}$, this is simply 
\begin{equation}
\domd = \left\{(t,\theta,\varphi)\left| \, |t| \leq \theta_{\cal A}- \theta, \theta \in (0,\theta_{\cal A})\right. \right\} \cup  \left\{(t,\theta,\varphi)\left| \, |t| \leq \theta_{\cal A}+ \theta, \theta \in (-\theta_{\cal A},0) \right.\right\}
\label{}
\end{equation}	
with future and past tips $q^\wedge = (t=\theta_{\cal A},\theta =0, \varphi=0)$ and 
$q^\vee = (t=-\theta_{\cal A},\theta =0, \varphi=0)$, respectively. Constructing the bulk causal wedge is then simply a matter of examining the null geodesics from these tips: the future boundary $\bcwedgef$ is generated by ingoing null geodesics from $q^\wedge$ while the past boundary $\bcwedgep$ is generated by ingoing null geodesics from $q^\vee$. The null surfaces generated by these intersect at the $t=0$ slice in the bulk. The simplest of these null generators is the radially ingoing geodesic; we can track this to ascertain how far into the bulk the surface $\csf{\cal A}$ reaches. Solving explicitly for the null geodesics in the geometry \eqref{eq:metricansatz} we find an implicit equation for $\rX$, the radial penetration depth:
\begin{equation}\label{eq:wedgedepth}
\theta_{\mathcal{A}} = \int_{\rX}^{\infty}dr\, \frac{e^{\beta(r)/2}}{g(r)} 
\end{equation}
We begin in \S\ref{sec:radialextent} with a discussion of how deep these surfaces reach, then move on in \S\ref{sec:CIsurface} to map out the full surface.  In \S\ref{sec:chi} we explicitly evaluate the causal holographic information $\chi_{\cal A}$.

\subsection{Radial extent of the causal wedge}
\label{sec:radialextent}

\begin{figure}[!htb]
\begin{center}
\vskip1em
\includegraphics[width=0.45\textwidth]{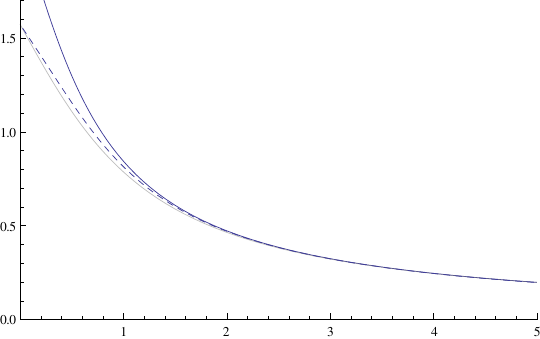}
\hskip1em
\includegraphics[width=0.45\textwidth]{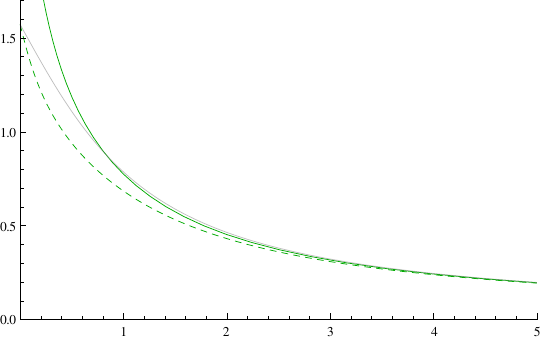}
\setlength{\unitlength}{0.1\columnwidth}
\begin{picture}(0.3,0.4)(0,0)
\put(-9.35,3.1){\makebox(0,0){$\theta_{\mathcal{A}}$}}
\put(-4.6,3.1){\makebox(0,0){$\theta_{\mathcal{A}}$}}
\put(0.4,0.15){\makebox(0,0){$\rE, \rX$}}
\end{picture}
\end{center}
\vskip-0.5em
\caption{Deepest extent of $\csf{\cal A}$ (solid lines) and $\extr{\cal A}$ (dashed lines) surfaces in soliton backgrounds, with AdS$_4$ shown in grey for comparison. Left: A $\Delta=2$ soliton in the phenomenological model. Right: A $\Delta=1$ soliton in the $U(1)^4$ truncation.  Note that the solid curves have finite $\theta_{\mathcal{A}}$-intercept (which we denote $\theta_{\mathcal A}^\text{tof}$), which is greater than $\frac{\pi}{2}$ in both cases.}
\label{fig:AllExtents}
\end{figure}
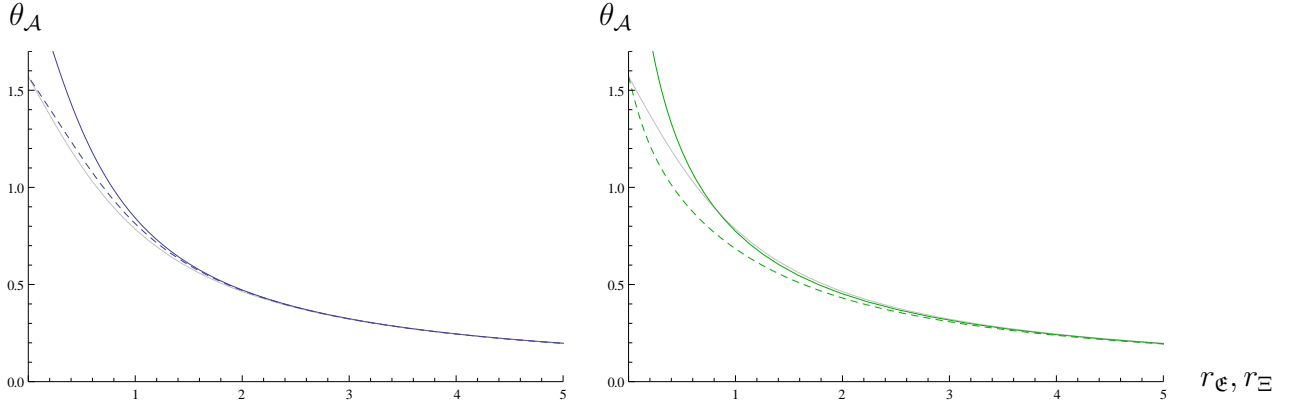

Recall that for global AdS spacetime the causal information surface coincides with the minimal surface hanging from the same $\partial\mathcal{A}$ (for polar-cap regions). As discussed in \cite{Hubeny:2012wa} this is not true for generic deformations of AdS, and indeed this is borne out by explicit computations.  In Fig.~\ref{fig:AllExtents} we plot $(\rX,\theta_{\mathcal{A}})$ curves for the soliton geometries studied so far, showing also where the minimal surfaces $\extr{\cal A}$ reach (i.e., $\rE$) for comparison. For very small regions $\theta_{\mathcal{A}} \ll 1$, the surfaces sit in the asymptotic region, leading to  $\rE=\rX=\rE^{\mathrm{AdS}}$; departures are however clear as we move to finite size regions. 

Once again there is a difference in the causal wedges depending on the choice of scalar boundary condition:
\begin{itemize}
\item For fixed $\theta_{\mathcal{A}}$, we find that $\rE^{\mathrm{AdS}}<\rE<\rX$ for $\Delta=2$ phenomenological solitons. 
\item  However, for regular $\Delta=1$ solitons in the $U(1)^4$ truncation we find
\begin{equation}
\begin{array}{cc}
\rE<\rX<\rE^{\mathrm{AdS}} & \textrm{for small } \theta_{\mathcal{A}} \\ 
\rE<\rE^{\mathrm{AdS}}<\rX & \textrm{for large } \theta_{\mathcal{A}} 
\end{array}
\end{equation}
In Fig.~\ref{fig:SubtractedExtents} we subtract off the AdS curve from the $(\rX,\theta_{\mathcal{A}})$ curves to illustrate this behaviour more clearly. 
\end{itemize}
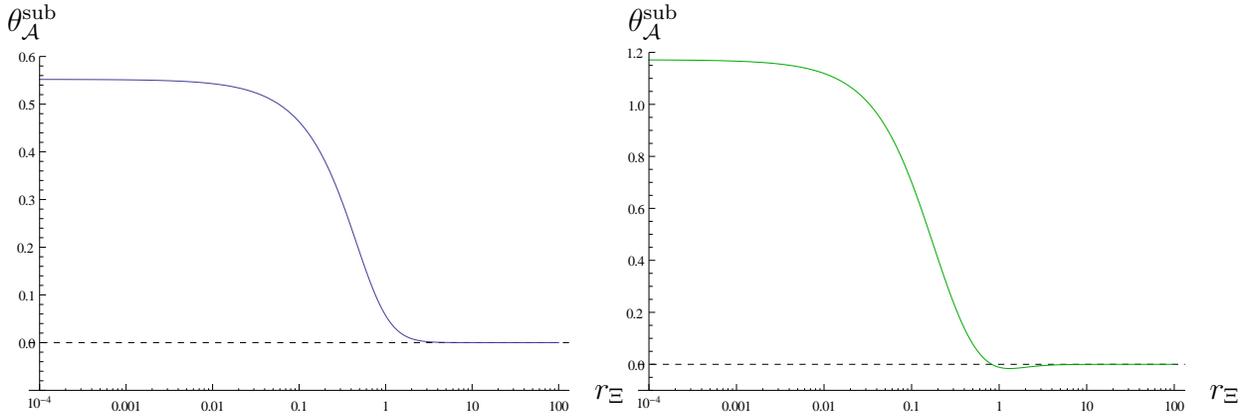
\begin{figure}[!htb]
\begin{center}
\vskip1em
\includegraphics[width=0.45\textwidth]{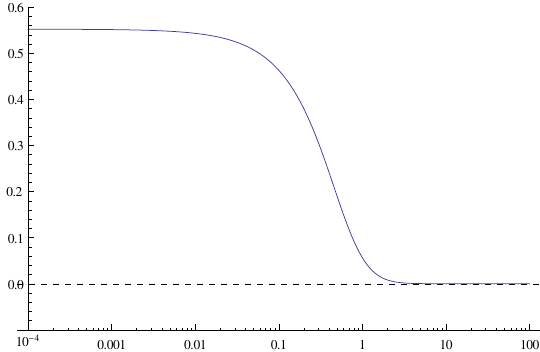}
\hskip1.5em
\includegraphics[width=0.45\textwidth]{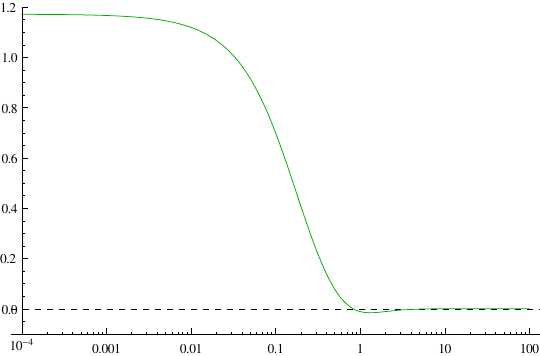}
\setlength{\unitlength}{0.1\columnwidth}
\begin{picture}(0.3,0.4)(0,0)
\put(-9.45,3.15){\makebox(0,0){$\theta_{\mathcal{A}}^{\mathrm{sub}}$}}
\put(-4.8,0.15){\makebox(0,0){$\rX$}}
\put(-4.45,3.15){\makebox(0,0){$\theta_{\mathcal{A}}^{\mathrm{sub}}$}}
\put(0.15,0.15){\makebox(0,0){$\rX$}}
\end{picture}
\end{center}
\vskip-0.5em
\caption{Deepest extent of $\csf{\cal A}$ in a regular soliton with $\Delta=2$ in the phenomenological model (left) and with $\Delta=1$ in the $U(1)^4$ truncation (right), with $\theta_{\mathcal{A}}^{\mathrm{sub}}\equiv \theta_{\mathcal{A}}-\cot^{-1}\rX$.}
\label{fig:SubtractedExtents}
\end{figure}

We recall that in general the minimal surfaces are required to lie outside the causal wedge for general deformations of \AdS{} \cite{Hubeny:2012wa, Wall:2012uf, Hubeny:2013gba}. This is clearly upheld, with $\rE < \rX$ irrespective of the choice of boundary condition. What is curious is that  for small regions in the $\vev{{\cal O}_1}$ coherent states the causal wedge is also attracted to the core region --- just like the minimal surfaces.

The attraction of the causal wedge into the bulk raises a potential question: can it be  that radial null geodesics travel through the bulk `faster' than along the boundary? If so we would encounter a serious causality violation in the CFT state. As discussed in  \cite{Hubeny:2006yu}, correlation functions of local operators on the boundary will generically have bulk-cone singularities whenever the operator insertion points are related by a null geodesic through the bulk. These singularities are however required to lie inside the boundary light-cone in sensible CFT states. Indeed the time-delay results in asymptotically AdS spacetimes of \cite{Woolgar:1994ar, Gao:2000ga} guarantee this and they simply rely on the matter satisfying the null energy condition, which is certainly upheld for the models \eqref{eq:generalaction} under consideration.

What saves the day is the repulsion of the causal wedge for large $\theta_{\cal A}$. In causally trivial spacetimes, the fastest communication through the bulk occurs for  anti-podal boundary points along radial bulk null geodesics (geodesics carrying angular momentum are effectively more timelike and thus slower). The time of flight along such a geodesic is 
\begin{equation}
\Delta t = 2 \int_{0}^{\infty}dr\, \frac{e^{\beta(r)/2}}{g(r)}  \equiv 2\, \theta_{\mathcal A}^\text{tof}
\end{equation}
which from \eqref{eq:wedgedepth} is simply twice the $\theta_{\mathcal{A}}$-intercept of a $(\rX,\theta_{\mathcal{A}})$ curve. This time of flight should be compared to communication along a boundary null geodesic which takes a time $\Delta t = \pi$. From the 
$\theta_{\mathcal{A}}$-intercepts in Figs.~\ref{fig:AllExtents} and \ref{fig:SubtractedExtents} this is indeed upheld and so the bulk geodesic gets time-delayed in the regular solitons relative to AdS.\footnote{ Recall that all null geodesics (both bulk and boundary) in vacuum AdS travel between anti-podal boundary points in time $\Delta t = \pi$ (in units where the AdS radius $L=1$).} 

Physically it is clear how this is achieved: while the gravitational potential measured by $g(r)$ is steeper near the boundary for the $\vev{{\cal O}_1}$ solitons, the null geodesic also has to contend with the red-shift factor measured by $e^{\beta(r)}$ whose effect is more pronounced near the core. In effect, while the null geodesics locally experience a speed up near the boundary of the soliton spacetime, they slow down sufficiently as they reach into the core region, ensuring that bulk causality remains consistent with boundary causality.

\subsection{Causal information surface $\csf{\cal A}$}
\label{sec:CIsurface}

Let us now turn to the causal wedge itself and examine whether $\csf{\cal A}$ is connected or disconnected. We focus on solitons of the phenomenological model for definiteness.

The null geodesic congruences are obtained by working in an effective three-dimensional geometry (exploiting the $U(1)$ isometry along $\partial_\varphi$) and satisfy
\begin{equation}\label{eq:wedgegeodesicaffine}
\dot{t}=\frac{e^{\beta(r)}}{g(r)}, \qquad \dot{r}^2 = -V_{\mathrm{eff}}(r) \equiv g(r) \left( \frac{e^{\beta(r)}}{g(r)} - \frac{\ell^2}{r^2} \right), \qquad \dot{\theta} = \frac{\ell}{r^2}
\end{equation}
where $\ell \in [0,1]$ is the (rescaled) angular momentum associated with motion in $\theta$ and a dot denotes differentiation with respect to the affine parameter. 
The potential $V_{\mathrm{eff}}(r)$ exhibits a centrifugal barrier at the origin (from the $\ell^2/r^2$ term) and 
asymptotes to $\ell^2 -1$ near the boundary of the spacetime.

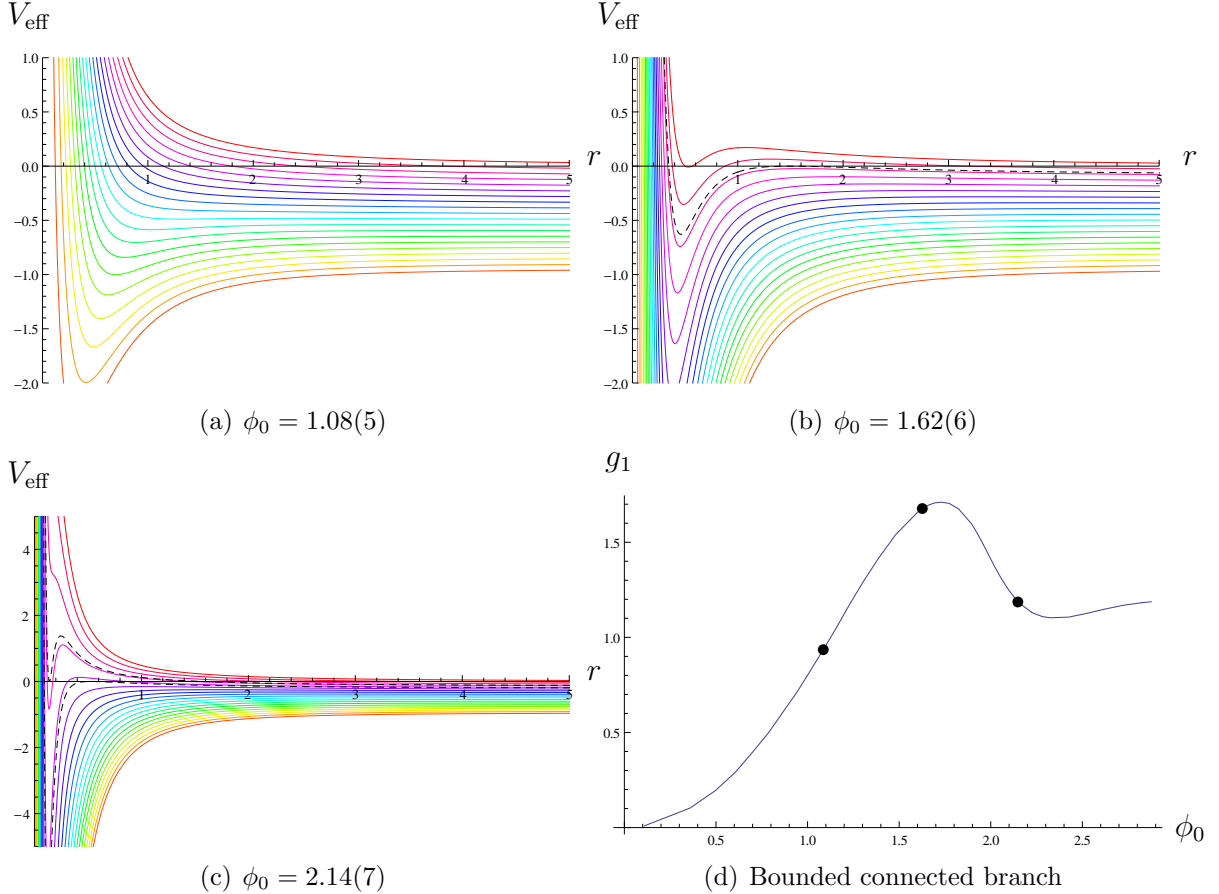
\begin{figure}[htb!]
\centering
\vskip1em
\subfigure[$\phi_0=1.08(5)$]{
\includegraphics[width=0.45\textwidth]{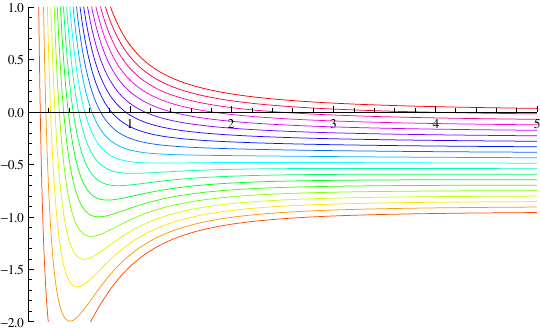}
}
\subfigure[$\phi_0=1.62(6)$]{
\includegraphics[width=0.45\textwidth]{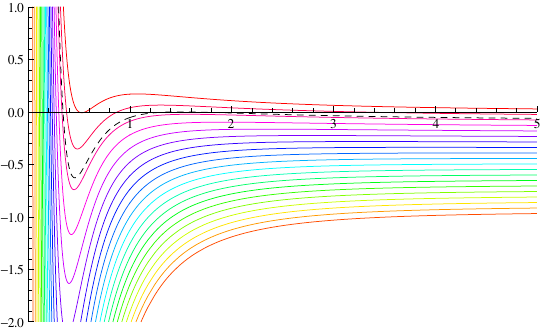}
}
\\ 
\vskip1em
\subfigure[$\phi_0=2.14(7)$]{      
\includegraphics[width=0.45\textwidth]{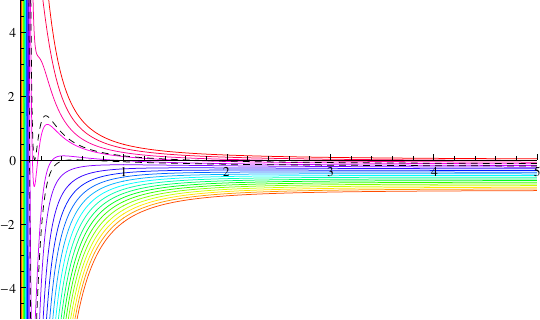}
}
\subfigure[Bounded connected branch]{
\includegraphics[width=0.45\textwidth]{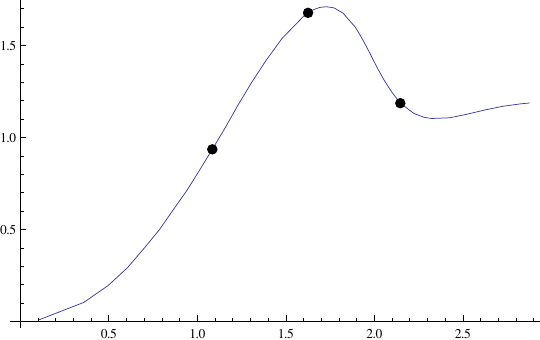}
\label{fig:EffectivePotentialTypeIbranch}
}
\begin{center}
\setlength{\unitlength}{0.1\columnwidth}
\begin{picture}(0.3,0.4)(0,0)
\put(-4.35,7.8){\makebox(0,0){$V_{\mathrm{eff}}$}}
\put(0.2,6.65){\makebox(0,0){$r$}}
\put(0.4,7.8){\makebox(0,0){$V_{\mathrm{eff}}$}}
\put(5.0,6.65){\makebox(0,0){$r$}}
\put(-4.35,4.1){\makebox(0,0){$V_{\mathrm{eff}}$}}
\put(0.2,2.5){\makebox(0,0){$r$}}
\put(0.4,4.2){\makebox(0,0){$g_1$}}
\put(5.0,1.25){\makebox(0,0){$\phi_0$}}
\end{picture}
\end{center}
\vskip-3.5em
\caption{Effective potentials for $\Delta=2$ solitons on the bounded connected branch in the phenomenological theory. We plot $\ell^2=0.05,0.1,0.15,\ldots,1$ from bottom to top.  Dashed black  curves denote values of $\ell^2$ for which null circular orbits exist: $\ell^2 = 0.914(0)$ in (b) and $\ell^2 = 0.786(5)$ or $0.858(8)$ in (c).  The final panel indicates the position of these three solitons along the branch.}
\label{fig:EffectivePotentialTypeI}
\end{figure}

The presence of the centrifugal barrier indicates that geodesics with $\ell \neq 0$ will always have a turning point for some $r >0$ where $V_{\mathrm{eff}}(r)$ has a simple zero. However, by tuning the core value of the scalar we can cause the geometry to have a null circular orbit. For this we require $V_{\mathrm{eff}}(r)$ to have a double zero and so satisfy
\begin{equation}
\frac{d}{dr}\left(\frac{e^{-\beta(r)}\, g(r)}{r^2}\right)\bigg|_{r=r_0}=0\,,\qquad \textrm{and} \quad \ell_0^2=\frac{r^2_0\, e^{\beta(r_0)}}{g(r_0)} \,,
\end{equation}
Note that the angular momentum on the circular orbit $\ell_0^2(\phi_0)$ has a maximum at $\phi_0^\textrm{max}$ for the bounded connected branch of solitons and is monotonically increasing for the unbounded branch. In Fig.~\ref{fig:EffectivePotentialTypeI} we plot $V_{\textrm{eff}}(r)$ for $\Delta=2$ solitons on the  bounded  connected branch in the phenomenological theory.\footnote{For the $\vev{{\cal O}_1}$ solitons of the $U(1)^4$ truncation we never encounter a circular orbit. Relatedly the casual wedge has a trivial topology for any $\theta_{\cal A}$. We will therefore refrain from describing the causal wedges in this case explicitly in \S\ref{sec:CIsurface} since the general features are comprehensively exhibited in the $\vev{{\cal O}_2}$ solitons of the phenomenological model.} 

Consider the  null geodesic congruences from $q^\wedge$, $q^\vee$. There are four distinct sets of these, labelled by their temporal and spatial orientations: $P$(past), $F$(future) and $L$(left) and $R$(right), respectively. These are further indexed by the conserved angular momentum. In particular, $PR_\ell$ and $PL_\ell$ generate $\bcwedgep$ while $FL_\ell$ and $FR_\ell$ generate $\bcwedgef$. The congruences intersect at $t=0$  along a spatial two-dimensional surface $\mathcal{X}_{t=0}$. 

As described in \cite{Hubeny:2013gba}, this surface is $\csf{\cal A}$ itself as long as it does not self-intersect.  In this case $\bcwedge$ is topologically trivial. The necessary condition for this to happen is that generators of $PL$ and $FL$ intersect each other (and similarly for $PR$ and $FR$) at $t=0$.

However, it could be that the $PR$ generators intersect the $PL$ generators for some $t<0$ (similarly, the $FL$ and $FR$ generators meet at some $t>0$). If this happens then  $\mathcal{X}_{t=0}$ self-intersects (as it does for large enough ${\cal A}$ in Schwarzschild-\AdS{}) and $\csf{\cal A}$ closes off at $\theta = \pi$, leaving a hole in $\cwedge$.  

Armed with the information of the turning points we can now integrate \eqref{eq:wedgegeodesicaffine} to find $\mathcal{X}_{t=0}$.  For each $\ell$ we first find the radial position $r_{t=0}(\ell)$ where $t_{\ell}(r)=0$ (which may occur on either side of  the turning point) then evaluate $\theta_{t=0}(\ell)$ at this radius.  If $\theta_{t=0}(\ell) > \pi$ then it must be that $PL_\ell$ and $PR_\ell$ already intersected at an earlier time and so these generators are no longer on the boundary of the casual wedge $\bcwedge$ (rather, they lie inside). This is the signal that $\csf{\cal A}$ is made of disconnected components. In all examples examined we encountered $\csf{\cal A}$ with at most two disconnected components.

It was further conjectured in \cite{Hubeny:2013gba} that a disconnected $\csf{\cal A}$ (in spherically-symmetric spacetimes as those discussed here) is correlated with the existence of a null circular orbit in the spacetime. We shall now present some explicit evidence in favour of this conjecture.

To understand this phenomenon and to determine the critical region size $\theta_{\cal A}^*$ where $\csf{\cal A}$ breaks up, it suffices to focus on the behaviour of $\theta_{t=0}(\ell)$. We have $\theta_{t=0}(1)=\theta_{\cal A}$ since $\ell =1$ geodesics stay on the boundary.  On the other hand, $\theta_{t=0}(0) \in \{0,\pi\}$ depending on whether the radially ingoing null geodesic crosses the origin. Recall from Fig.~\ref{fig:AllExtents} that the null geodesic from $q^\wedge_\text{tof} = \left(t=\theta_{\mathcal{A}}^{\mathrm{tof}},\theta_{\mathcal A}= 0 \right)$ on the boundary makes it to the origin ($r=0$) at $t=0$. This means that radial null geodesics with $\theta_{\cal A}< \theta_{\mathcal{A}}^{\mathrm{tof}}$ intersect the $t=0$ plane on the northern hemisphere and for larger regions they cross over the equator to the southern hemisphere of the ${\bf S}^2$ (we take the mid-point $\theta = 0$ of ${\cal A}$ to be the north pole). To wit,
\begin{equation}
\theta_{t=0}(0) =
\begin{cases}
& 0	\,, \qquad \theta_{\cal A} \leq \theta_{\mathcal{A}}^{\mathrm{tof}}\\
& \pi \,, \qquad \theta_{\cal A} > \theta_{\mathcal{A}}^{\mathrm{tof}}
\end{cases}	
\label{}
\end{equation}	
 The curve $\theta_{t=0}(\ell)$ then connects these two end-points, but it can do so in an interesting manner. Let us consider the two cases discussed above in turn.

\paragraph{(i) $\theta_{\cal A} \leq \theta_{\mathcal{A}}^{\mathrm{tof}}$:}
Consider first the case when $\theta_{t=0}(0) = 0$, so that the radial null geodesic intersects the $t=0$ plane on the northern hemisphere of the ${\bf S}^2$. Then while for small $\theta_{\cal A}$ we will see $\theta_{t=0}(\ell)$ monontonically increasing in $(0,\theta_{\cal A})$, this should cease to hold for larger regions.  This will happen the moment the geodesics start to wrap around and so $\theta_{t=0}(\ell)$ will develop a characteristic maximum $\ell_{\mathrm{max}}$. The geodesics with this value of the angular momentum are such that they turn around at $t=0$; the curve $t_{\ell_{\mathrm{max}}}(r)$ is reflection symmetric. As argued above, the causal information surface $\csf{\cal A}$ will have a single connected component as long as $\theta_{t=0}(\ell)$ does not cross $\pi$ and so the necessary condition for this to hold is simply
$\theta_{t=0}(\ell_{\mathrm{max}}) < \pi$.
  
\begin{figure}[!hp]
\centering
\subfigure{
\includegraphics[width=0.35\textwidth]{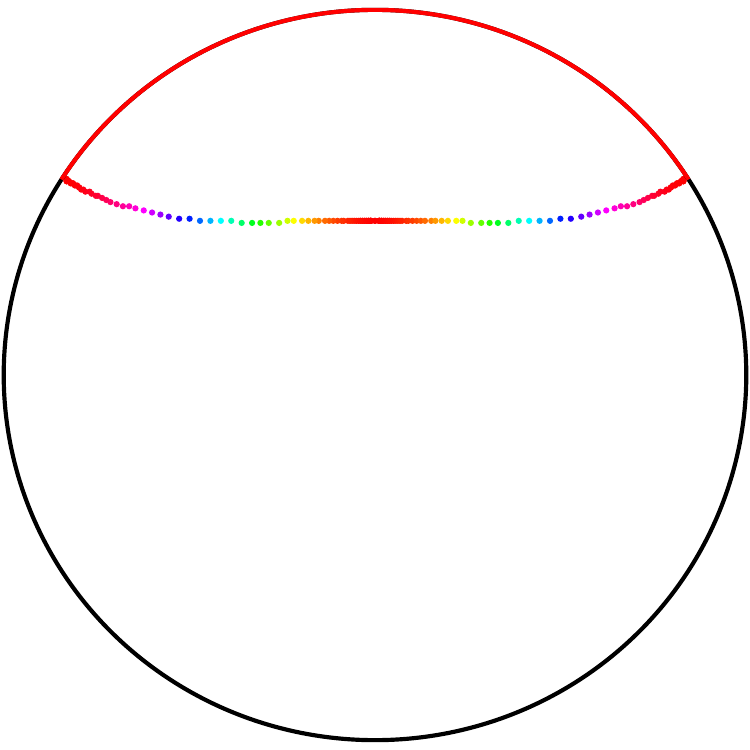}
}
\subfigure{
\includegraphics[width=0.5\textwidth]{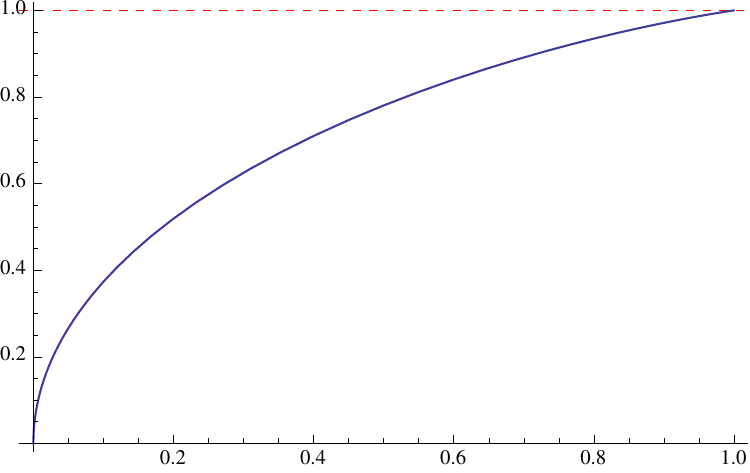}
}
\\ 
\vskip1em
\subfigure{      
\includegraphics[width=0.35\textwidth]{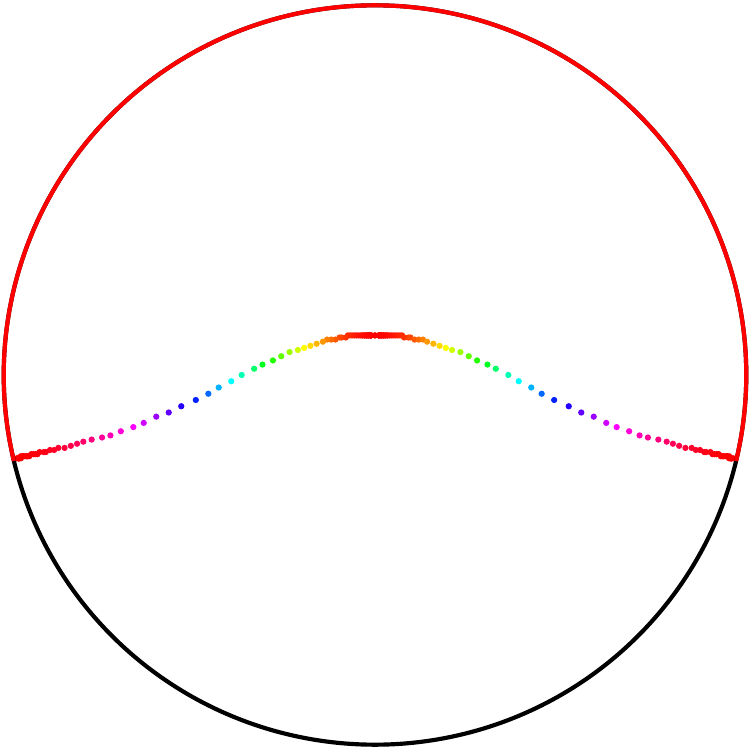}
}
\subfigure{
\includegraphics[width=0.5\textwidth]{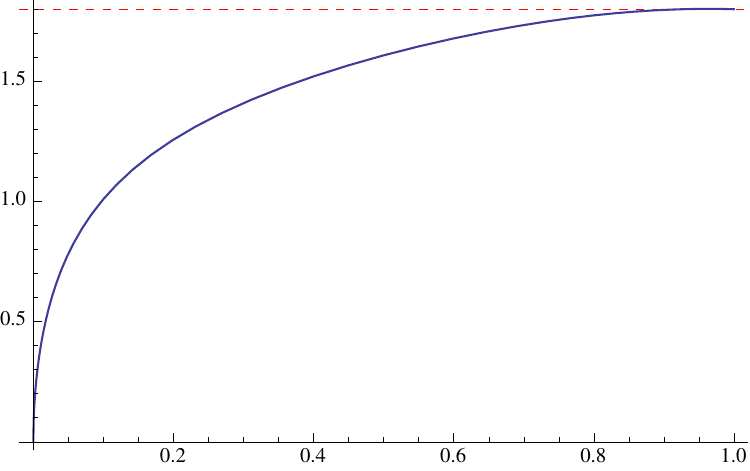}
}
\begin{center}
\setlength{\unitlength}{0.1\columnwidth}
\begin{picture}(0.3,0.4)(0,0)
\put(-0.3,8.1){\makebox(0,0){$\theta_{t=0}$}}
\put(4.95,4.9){\makebox(0,0){$\ell^2$}}
\put(-0.3,4.25){\makebox(0,0){$\theta_{t=0}$}}
\put(4.95,1.){\makebox(0,0){$\ell^2$}}
\put(-2.45,6.7){\makebox(0,0){${\cal X}_{t=0}=\color{blue}{\csf{\cal A}}$}}
\put(-4.0,7.8){\makebox(0,0){$\color{red}{{\cal A}}$}}
\put(-2.45,2.1){\makebox(0,0){${\cal X}_{t=0}=\color{blue}{\csf{\cal A}}$}}
\put(-4.0,3.95){\makebox(0,0){$\color{red}{{\cal A}}$}}
\end{picture}
\end{center}
\vskip-3.5em
\caption{Plots to show connected $\csf{\cal A}$ in a $\Delta=2$ bounded connected soliton with $\phi_0=1.08(5)$ in the phenomenological theory (far left dot in Fig.~\ref{fig:EffectivePotentialTypeIbranch}).  Two different values of $\theta_{\mathcal{A}}$ are shown (top row, $1$ and bottom row, $1.8$), both of which are below $\theta_{\mathcal{A}}^{\mathrm{tof}}=2.12(2)$.  Black dots denote values of $\rX$.}
\label{fig:WedgesWithoutHoles}
\end{figure}

In Fig.~\ref{fig:WedgesWithoutHoles} we exhibit examples of this behaviour $\csf{\cal A}$ for various $\theta_{\cal A} \leq \theta_{\mathcal{A}}^{\mathrm{tof}}$ in a $\Delta=2$ soliton on the bounded connected branch in the phenomenological theory. In neither case does the geometry admit  a null circular orbit and the causal information surface is indeed composed of a single connected component.\footnote{ We show the projection of the causal wedge boundary on the spatial $t=0$ slice. The behaviour of the full causal wedge can be intuited from Fig.~3 of \cite{Hubeny:2013gba}. In the soliton geometries we don't have a black hole horizon, but the operative feature is as we have emphasized at several times, the deep gravitational well in the core region.}

The causal information surface starts to pinch off as soon as $\theta_{t=0}(\ell_{\mathrm{max}}) = \pi$. This clearly has to happen for a critical region size $\theta_{\cal A}= \theta_{\cal A}^*$.  For $\theta_{\cal A} > \theta_{\cal A}^*$ the causal information surface $\csf{\cal A}$ has two distinct components.  The generators of the two segments are demarcated by two solutions of $\theta_{t=0}(\ell)=\pi$, labelled by $0<\ell_1<\ell_{\mathrm{max}}<\ell_2<1$.\footnote{ Note that these $\ell_{1,2}$ and $\ell_{\mathrm{max}}$ are all distinct from the value(s) of $\ell$ at which a null circular orbit exists in the soliton.}  One part is connected to the boundary and is given by $\mathcal{X}_{t=0}$ for $\ell\in(\ell_2,1)$. The other part is disconnected from the boundary and wraps the soliton core and is given by $\mathcal{X}_{t=0}$ for $\ell\in(0,\ell_1)$.  

\begin{figure}[!tp]
\begin{center}
\includegraphics[width=0.35\textwidth]{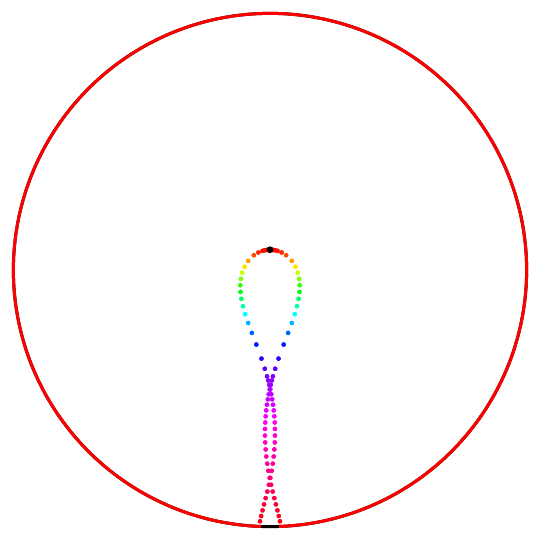}
\hskip1em
\includegraphics[width=0.5\textwidth]{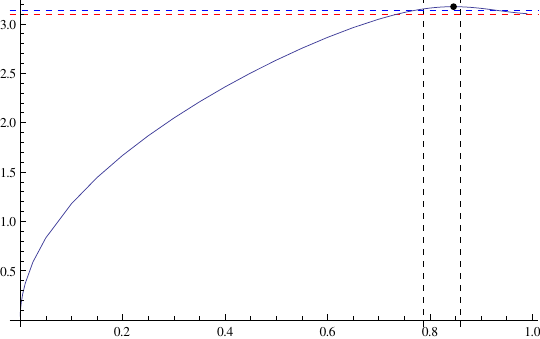}
\setlength{\unitlength}{0.1\columnwidth}
\begin{picture}(0.3,0.4)(0,0)
\put(-4.95,3.4){\makebox(0,0){$\theta_{t=0}$}}
\put(0.15,0.15){\makebox(0,0){$\ell^2$}}
\put(-8,1.1){\makebox(0,0){${\cal X}_{t=0}$}}
\put(-6.45,1.1){\makebox(0,0){$\color{blue}{\csf{\cal A}}$}}
\put(-6.7,1.1){\vector(-1,1){0.25}}
\put(-6.7,1.1){\vector(-1,-2){0.4}}
\put(-8.75,3.2){\makebox(0,0){$\color{red}{{\cal A}}$}}
\end{picture}
\end{center}
\vskip-0.5em
\caption{Plots to show the disconnected components of $\csf{\cal A}$ in a $\Delta=2$ bounded connected soliton with $\phi_0=2.14(7)$ in the phenomenological theory (far right dot in Fig.~\ref{fig:EffectivePotentialTypeIbranch}).  Left: $\mathcal{X}_{t=0}$, composed of individual intersection points colour-coded by $\ell$. The thick red arc is the spatial boundary region $\mathcal{A}$ with $\theta_{\mathcal{A}}=3.1$ and the black dot corresponds to $\rX=0.124(5)$ ($\ell=0$). Right: $\theta_{t=0}$ as a  function of $\ell^2$. The horizontal dashed blue  line is at $\pi$ and the horizontal dashed red   line is at $\theta_{\mathcal{A}}=3.1$, whereas the  vertical black dashed lines indicate null circular orbits at $\ell^2_0 = 0.786(5)$ and $0.858(8)$. The black dot denotes the maximum, which is found as described in the text. Note that $\theta_{\mathcal{A}}^{\mathrm{tof}} = 4.79(8)$ for this soliton.}
\label{fig:WedgeHole}
\end{figure}

In Fig.~\ref{fig:WedgeHole} we exhibit a disconnected $\Xi_{\mathcal{A}}$ in a particular $\vev{{\cal O}_2}$ soliton  on the bounded connected branch in the phenomenological theory that admits a null circular orbit.   Note that the spatial boundary surface $\mathcal{A}$ must be  quite large in order to see this phenomenon; for the soliton shown, the critical size above which $\csf{\cal A}$  disconnects is $\theta_{\mathcal{A}}^*\simeq 3.07$.  In all cases we checked, this break-up of $\csf{\cal A}$ is correlated with the presence of a circular orbit. 

Note that at the critical value $\theta_{\cal A}^*$, one necessarily has $\ell_1=\ell_2=\ell_{\mathrm{max}}$. As in \cite{Hubeny:2013gta} we denote this critical value of $\ell_{\mathrm{max}}$ by $\ell_*$. Geodesics with $\ell=\ell_*$ are special: they 
smoothly connect the tips $q^\vee$ and $q^\wedge$ of $\domd$, turning around symmetrically at $t=0$.\footnote{ Note in particular that $\ell_*^2 = 0.84(7) \neq \ell^2_0$ for either  $\ell_0$ quoted in the caption of Fig.~\ref{fig:WedgeHole}. In \cite{Hubeny:2013gba} it was indicated that in Schwarzschild-\AdS{d+1} geometries, $\ell_* \approx \ell_0$ (the deviation was $O(10^{-3})$ and was in fact used to determine $\theta_{\cal A}$. This appears to have been an curious coincidence and doesn't seem to extend to the soliton examples.}

\paragraph{(ii) $\theta_{\cal A} > \theta_{\mathcal{A}}^{\mathrm{tof}}$:} When the radial null geodesic intersects the $t=0$ plane in the southern hemisphere, then $\theta_{t=0}(\ell)$ starts out at $\pi$ and has to eventually get down to $\theta_{\cal A} < \pi$. 
For the case that $\theta_{t=0}(\ell)\in[\theta_{\cal A},\pi]$, again there is a  single connected component of $\csf{\cal A}$ that coincides with ${\cal X}_{t=0}$, this time lying entirely in the southern hemisphere. 

However, it could be that geodesics with $\ell \approx 0$ attain $\theta_{t=0} > \pi$, whence the curve would remain above $\pi$, peak at some intermediate $\ell_{\mathrm{max}}$ before descending back to $\theta_{\cal A}$. Effectively, we would have 
$\ell_1 = 0 < \ell_{\mathrm{max}} < \ell_2 < 1$. Now the parts of the curve ${\cal X}_{t=0}$ for $\ell \leq \ell_2$ are no longer on the boundary of the causal wedge. As before these generators enter into the bulk of the causal wedge, the components $PL_\ell$ and $PR_{\ell}$  (likewise $FR_{\ell}$ and $FL_{\ell}$) having met below (above) $t=0$. This segment of ${\cal X}_{t=0}$ simply represents a curve of caustics. The  piece of ${\cal X}_{t=0}$ for $\ell > \ell_2$ generates the causal information surface $\csf{\cal A}$, which we emphasise has a single connected component.

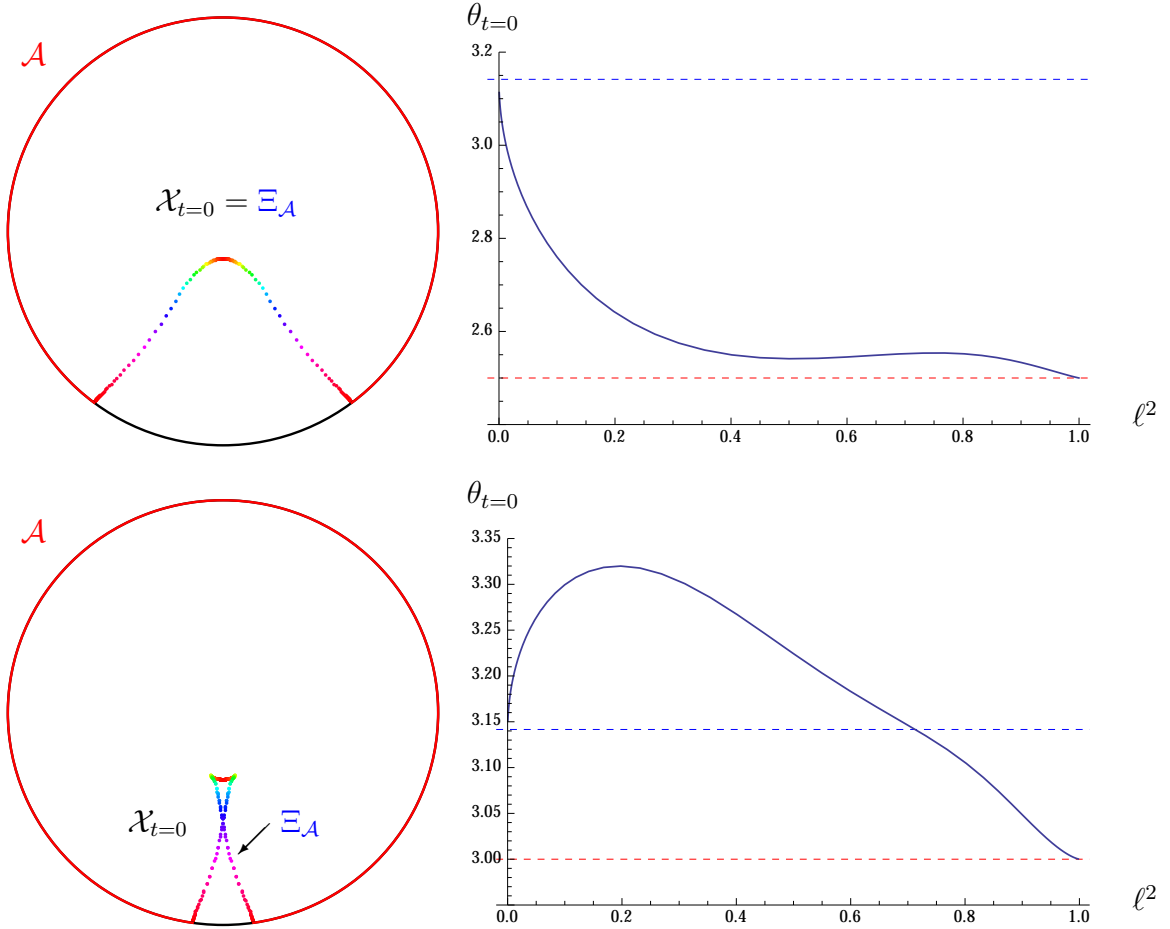
\begin{figure}[!hp]
\centering
\subfigure{      
\includegraphics[width=0.35\textwidth]{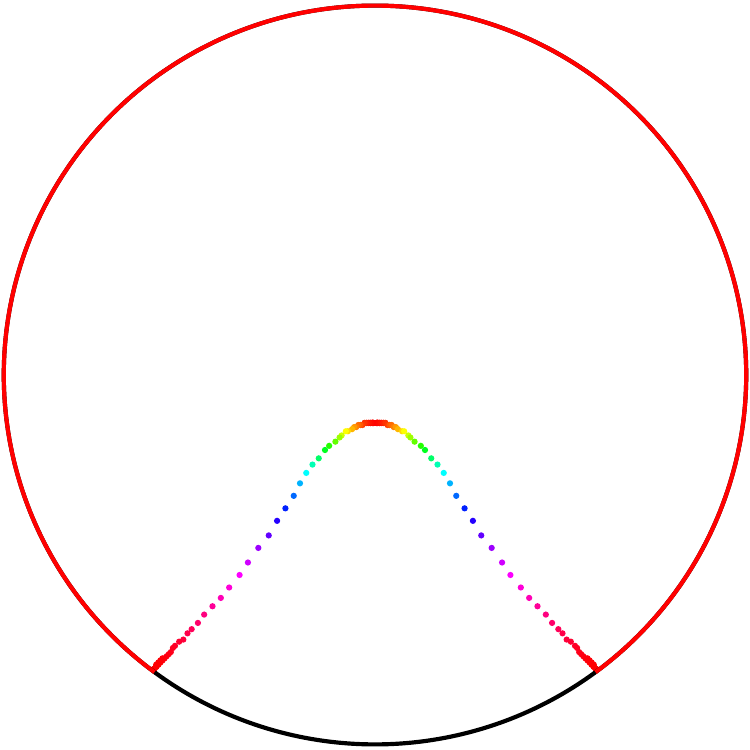}
}
\subfigure{
\includegraphics[width=0.5\textwidth]{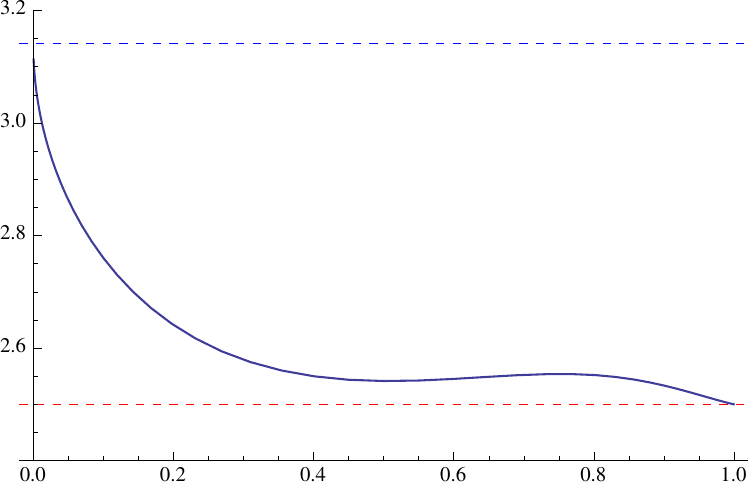}
}
\\ 
\vskip1em
\subfigure{      
\includegraphics[width=0.35\textwidth]{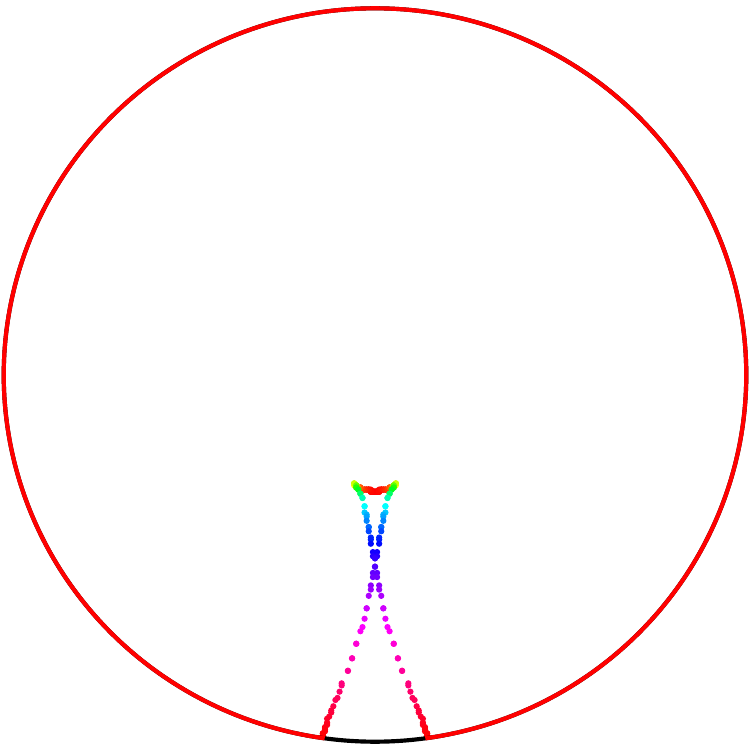}
}
\subfigure{
\includegraphics[width=0.5\textwidth]{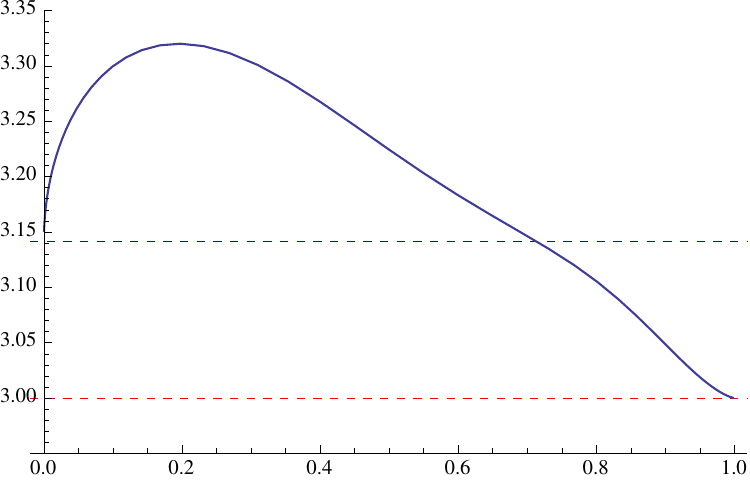}
}
\begin{center}
\setlength{\unitlength}{0.1\columnwidth}
\begin{picture}(0.3,0.4)(0,0)
\put(-0.3,8.1){\makebox(0,0){$\theta_{t=0}$}}
\put(4.95,4.9){\makebox(0,0){$\ell^2$}}
\put(-0.3,4.25){\makebox(0,0){$\theta_{t=0}$}}
\put(4.95,1.){\makebox(0,0){$\ell^2$}}
\put(-2.45,6.6){\makebox(0,0){${\cal X}_{t=0}=\color{blue}{\csf{\cal A}}$}}
\put(-4.0,7.8){\makebox(0,0){$\color{red}{{\cal A}}$}}
\put(-3.,1.6){\makebox(0,0){${\cal X}_{t=0}$}}
\put(-1.85,1.6){\makebox(0,0){$\color{blue}{\csf{\cal A}}$}}
\put(-2.1,1.6){\vector(-1,-1){0.25}}
\put(-4.0,3.95){\makebox(0,0){$\color{red}{{\cal A}}$}}
\end{picture}
\end{center}
\vskip-3.5em
\caption{Plots to show connected $\csf{\cal A}$ in a $\Delta=2$ bounded connected soliton with $\phi_0=1.08(5)$ in the phenomenological theory (far left dot in Fig.~\ref{fig:EffectivePotentialTypeIbranch}).  Two different values of $\theta_{\mathcal{A}}$ are shown (top row, $2.5$ and bottom row, $3$), both of which are above $\theta_{\mathcal{A}}^{\mathrm{tof}}=2.12(2)$.}
\label{fig:WedgesWithoutHoles2}
\end{figure}

In Fig.~\ref{fig:WedgesWithoutHoles2} we exhibit examples of this behaviour $\csf{\cal A}$ for various $\theta_{\cal A} \geq \theta_{\mathcal{A}}^{\mathrm{tof}}$ in a $\Delta=2$ soliton on the bounded connected branch in the phenomenological theory. Once again in neither case does the geometry admit  a null circular orbit.

\subsection{Causal holographic information $\chi_{\cal A}$}
\label{sec:chi}

We now turn to the computation of the causal holographic information associated with polar-cap regions of our soliton coherent states.  As described in \S\ref{sec:review} we will focus on the finite quantity 
$\Delta\chi_{\cal A}$ defined in \eqref{delchiA} which makes use of the knowledge that $\chi_{\cal A}^\text{vacuum} = S_{\cal A}^\text{vacuum}$ for polar-cap regions \cite{Hubeny:2012wa}.

\begin{figure}[h!]
\vskip2em
\begin{center}
\includegraphics[width=0.45\textwidth]{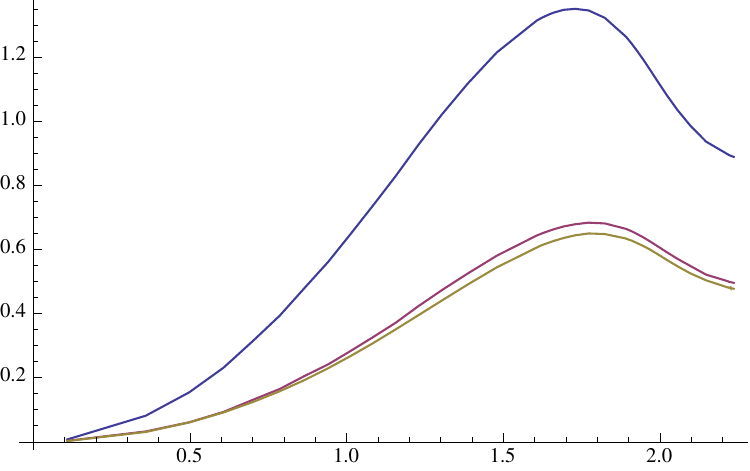}
\hskip1em
\includegraphics[width=0.45\textwidth]{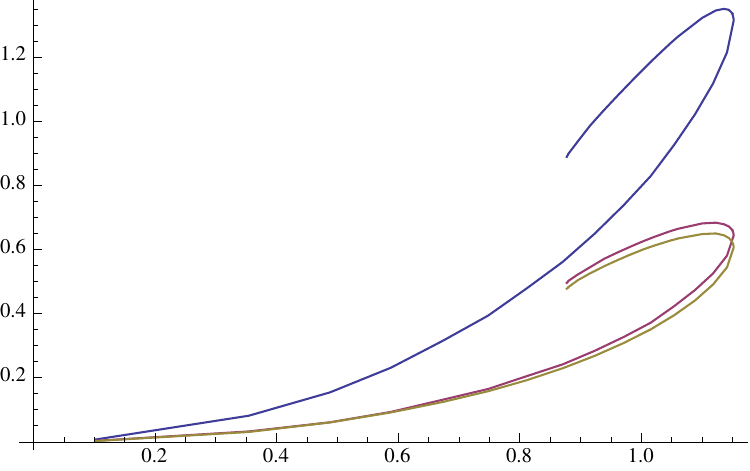}
\setlength{\unitlength}{0.1\columnwidth}
\begin{picture}(0.3,0.4)(0,0)
\put(-8.5,3.15){\makebox(0,0){$\{\Delta \hat{H}_{\cal A},\Delta \hat{\chi}_{\cal A},\Delta \hat{S}_{\cal A}$\}}}
\put(-3.7,3.15){\makebox(0,0){$\{\Delta \hat{H}_{\cal A},\Delta \hat{\chi}_{\cal A},\Delta \hat{S}_{\cal A}$\}}}
\put(-4.75,0.15){\makebox(0,0){$\phi_0$}}
\put(0.2,0.15){\makebox(0,0){$\vev{{\cal O}_2}$}}
\end{picture}
\end{center}
\vskip-0.5em
\caption{Comparison between the modular Hamiltonian, causal holographic information and entanglement entropy (rescaled) for $\Delta=2$ solitons on the bounded connected branch in the phenomenological model for $\theta_{\mathcal{A}}=1$. Curves are labelled from top to bottom.}
\label{fig:chidelta2}
\end{figure}

\begin{figure}[h!]
\vskip1.5em
\begin{center}
\includegraphics[width=0.45\textwidth]{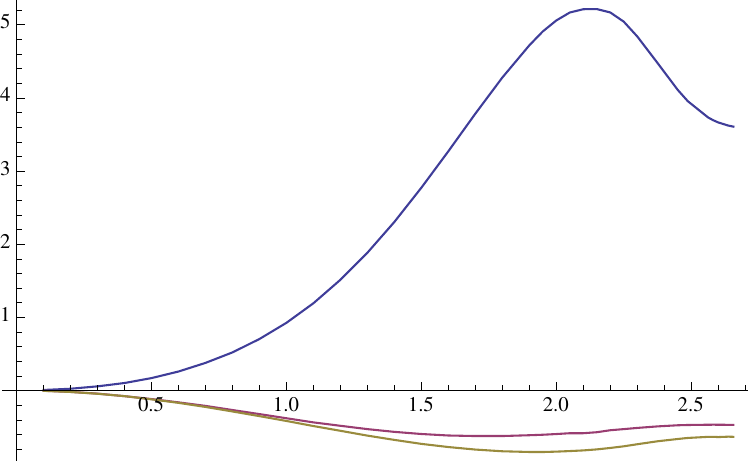}
\hskip1em
\includegraphics[width=0.45\textwidth]{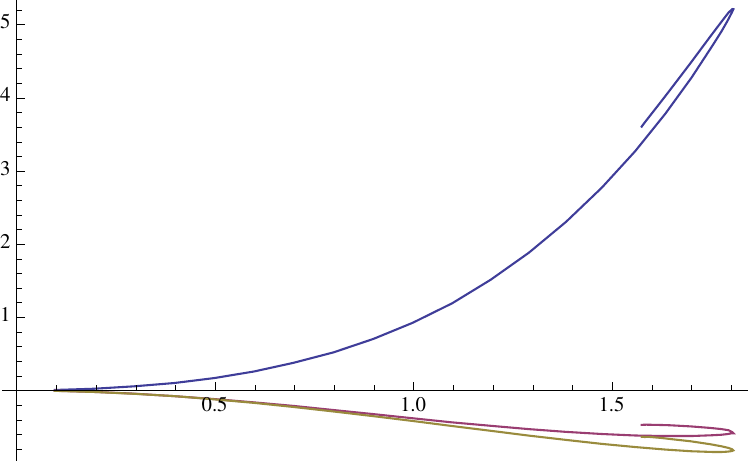}
\setlength{\unitlength}{0.1\columnwidth}
\begin{picture}(0.3,0.4)(0,0)
\put(-8.5,3.15){\makebox(0,0){$\{\Delta \hat{H}_{\cal A},\Delta \hat{\chi}_{\cal A},\Delta \hat{S}_{\cal A}$\}}}
\put(-3.8,3.15){\makebox(0,0){$\{\Delta \hat{H}_{\cal A},\Delta \hat{\chi}_{\cal A},\Delta \hat{S}_{\cal A}$\}}}
\put(-4.75,0.15){\makebox(0,0){$\phi_0$}}
\put(0.2,0.15){\makebox(0,0){$\vev{{\cal O}_1}$}}
\end{picture}
\end{center}
\vskip-1em
\caption{Comparison between the modular Hamiltonian, causal holographic information and entanglement entropy (rescaled) for $\Delta=1$ solitons on the bounded connected branch in the phenomenological model for $\theta_{\mathcal{A}}=1$. Curves are labelled from top to bottom.}
\label{fig:chidelta1}
\end{figure}
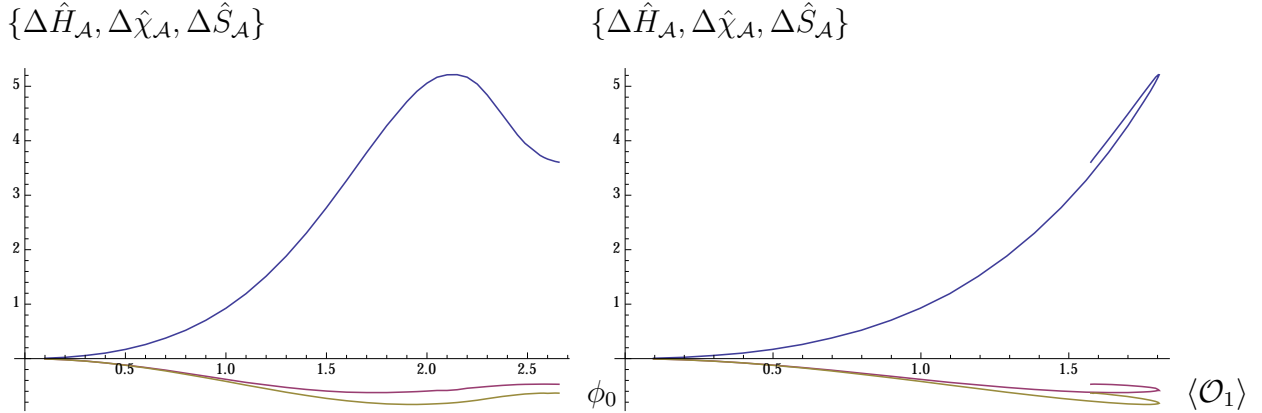

The area of $\csf{\cal A}$ can be computed as in \eqref{eq:areaequation} with the parameter choice $s=\ell^2$.  We focus on a region for which the causal information surface does not disconnect, i.e., $\theta_{\cal A} < \theta_{\cal A}^*$. Our results are presented in Figs.~\ref{fig:chidelta2} and \ref{fig:chidelta1} for $\vev{{\cal O}_2}$ and $\vev{{\cal O}_1}$ solitons in the phenomenological model, respectively.  

As previously anticipated on general grounds we find $\Delta H_{\cal A} > \Delta \chi_{\cal A} > \Delta S_{\cal A}$ for both quantizations.  We have already provided independent arguments for $\Delta H_{\cal A} > \Delta S_{\cal A}$ using the relative entropy and $\Delta \chi_{\cal A} > \Delta S_{\cal A}$
follows because the extremal surface is forced to lie outside the causal wedge 
\cite{Hubeny:2012wa, Wall:2012uf, Hubeny:2013gba} (for spacetimes satisfying the null energy condition).  It is not clear if $\Delta H_{\cal A}> \Delta \chi_{\cal A}$ is necessary in general.

Note that $\Delta \chi_{\cal A}$ is also negative for the $\Delta =1$ solitions. Once again this is easy to understand geometrically since the causal wedges get further away from the boundary due to the asymptotic speed-up of null geodesics discussed earlier. Also, since  this quantity is defined purely geometrically (see \cite{Hubeny:2012wa}), it is clear that there should be no scalar contribution to $\chi_{\cal A}$ and thus our subtraction scheme is  sensible.
 While we do not have much intuition for this quantity, our  results should serve as a constraint for any proposal, such as the one of \cite{Kelly:2013aja}, to understand it from the field theory perspective.

\section{Discussion}\label{sec:discussion}%

The thrust of the exploration in this paper has been to understand the nature of (holographic) information content in a class of bosonic coherent states in the boundary field theory. We described the behaviour of minimal surfaces and  causal wedges in two particular Einstein-Maxwell-scalar models: a phenomenological model with quadratic scalar interactions and a $U(1)^4$ truncation of eleven-dimensional supergravity. In each case we were allowed a choice of scalar field boundary condition; correspondingly we had states carrying  expectation values for a $\Delta=1$ operator ${\cal O}_1$ or a $\Delta =2$ operator ${\cal O}_2$.

The analysis of entanglement entropy for polar-cap regions ${\cal A}$ in the ${\cal O}_2$ solitons reflected vindication of conventional wisdom. Creating a bosonic coherent state by exciting modes of ${\cal O}_2$ not only increases the energy of the state but also results in an increased entanglement between the fundamental degrees of freedom (relative to the vacuum). Gravitationally this is easy to understand since the macroscopic population of the dual scalar field eigenmode in the bulk results in an increased gravitational potential in the core region that in turn repels the minimal surfaces.

On the other hand, for the $\vev{{\cal O}_1}$ solitons we observed the surprising phenomenon of entanglement reduction. While the coherent state has greater energy than the vacuum, it has less entanglement between the fundamental modes. In the bulk this has roots in the increased gravitational potential near the asymptotic region owing to the slow fall-off of the scalar field. From the behavior of the minimal surfaces we have seen that the competition between the core and asymptotic potentials can be subtle. However, the asymptotic potential dominates the computation of the entanglement entropy.

Our explicit results were obtained for operators with dimensions $\Delta_- =1$ (or bulk scalars of mass $m^2_\phi  = -2$) in some particular models, but the general result of entanglement reduction persists whenever we implement Neumann boundary conditions. Consider a bulk scalar field of mass in the window where the both scalar boundary conditions are allowed, i.e., $m^2_{BF} \leq m^2_\phi \leq m^2_{BF} +1$ with the Breitenlohner-Freedman bound $m^2_{BF} = -\frac{d^2}{4}$ in \AdS{d+1}. Expanding the perturbative result for $\vev{{\cal O}_{\Delta}}=\varepsilon$ from Appendix~\ref{app:pertubative} near the boundary we find
\begin{equation}
g(r) = r^2+1+\left[\frac{\Delta}{2}\,    r^{2 (1-\Delta )}-\frac{\sqrt{\pi } \Delta   \Gamma \left(\Delta -\frac{1}{2}\right)}{4  \Gamma (\Delta )}\frac{1}{r} + \ldots \right]\varepsilon^2 + O(\varepsilon^4)\quad \mathrm{for}\quad \Delta>1/2
\end{equation}
to lowest non-trivial order in $\varepsilon$.  Note the $r^{2 (1-\Delta )}$ term at $O(\varepsilon^2)$.  This appears before the $O(r^{-1})$ term if we impose alternate boundary conditions, i.e., for  $\Delta=\Delta_-\in\left(\frac{1}{2},\frac{3}{2}\right)$. Using the full $O(\varepsilon^2)$ result for general $\cal A$ we find
\begin{align}
\frac{1}{4\pi\, c_\text{eff}}\,\Delta S_{\cal A}
&= \frac{\pi \Delta (2 \Delta-3 )}{6}\int_{\cot\theta_{\cal A}}^\infty dr\, \frac{ r\left(r^2-\cot^2\theta_{\cal A}\right) \, _2F_1\left(\frac{3}{2},\Delta ;\frac{5}{2};-r^2\right)\sin\theta_{\cal A}}{ \left(r^2+1\right)^{3/2}}\, \varepsilon^2 \nonumber\\
&\phantom{=\ } + O(\varepsilon^4) 
\end{align}
%
%
The integrand is positive on $[\cot\theta_{\cal A},\infty)$ and so the entanglement entropy is indeed lower than for the vacuum when $\Delta\in\left(\frac{1}{2},\frac{3}{2}\right)$.\footnote{We note in passing that  at the lower boundary of this window, $\Delta=\frac{1}{2}$, where the unitary bound is saturated for scalar operators  we have 
$$
g(r) = r^2+\frac{r }{4}\, \varepsilon ^2+1-\frac{2 \log 2r-1}{8 r}\, \varepsilon ^2 + O(\varepsilon^4)
$$
instead. For this case the area integral does not converge. However, we think this is not a problem: there are no known examples of such operators in holographic models, reflecting the common lore that interacting QFTs do not admit operators saturating the unitarity bound. At the upper boundary, $\Delta=\frac{3}{2}$, we encounter  the Breitenlohner-Freedman bound, for which we must go to one higher  order in perturbation theory to see a change in the geometry.} 

We now wish to argue that the lowering of entanglement relative to the vacuum for Neumann boundary conditions is not an artifact of our having mis-identified the entanglement entropy in these geometries. \emph{A priori} one could image that there is a boundary term involving the slow fall-off mode  ($\phi_1$ in the $\Delta =1$ example) which should be included in the computation of the entanglement entropy.  A natural candidate is the scalar counter-term which is present for these boundary conditions and makes its presence felt in the computation of the boundary stress tensor. However, such a term is not present in the entanglement entropy prescription. Not only would it spoil the aesthetic beauty of the minimal area prescription of \cite{Ryu:2006bv}, but one can use the generalized gravitational entropy construction of \cite{Lewkowycz:2013nqa} to show that such counter-terms do not enter the computation of any of the R\'{e}nyi entropies (and thus by analytic continuation the entanglement entropy). The argument is straightforward: the divergence structure encountered while computing the replicated partition function cancels against the normalization of the reduced density matrix.

While the physical picture seems clear from the bulk, it is  not clear how the reduction of entanglement is achieved in the field theory directly. \emph{A priori} in a given CFT one expects there to be universal democracy in the space of relevant operators. There is no \emph{a priori} reason to single out (single-trace) operators ${\cal O}_\Delta$  whose multi-traces are also relevant (generically this requires $\Delta \leq \frac{d}{2}$, the regime of alternate boundary conditions). Clearly, the unconventional nature of entanglement in these coherent states should be understood from a field theory perspective.

It is also interesting to ask whether this phenomenon persists for extremal surfaces; despite the background state being static, these would be relevant if the region ${\cal A}$ we choose is on a boundary Cauchy surface that is not aligned with with the Killing field $\partial_t$. For these extremal surfaces we need to worry about the gravitational red-shift as well (just as for the causal wedges). We believe that the entanglement relative to the vacuum would be negative for small boundary regions, though there might be some turn-around for larger regions. 

We believe the issues we discussed here have a bearing on the recent ideas on trying to reconstruct Einstein's equations  from the entanglement entropy, especially the relation $\Delta S_{\cal A} = \Delta H_{\cal A}$ \cite{Blanco:2013joa, Lashkari:2013koa}. The fact that the entanglement entropy is reduced at the non-linear level (recall that $\Delta S_{\cal A}= \Delta H_{\cal A} =0$ in our soliton to leading order in $\vev{{\cal O}_\Delta}$) should indicate non-trivial constraints to this reconstruction programme at higher orders. To date the proposals exploit the linearized relations in simple states of the CFT carrying non-trivial $\vev{T^{\mu}_{\phantom{\mu}\nu}}$ with no other one-point functions non-vanishing; the dual geometries are thus solutions to vacuum Einstein's equations with a negative cosmological constant. One naively might have expected that the entanglement relations would allow recovery of Einstein's equations with a conserved bulk stress energy-momentum tensor (without particularly elucidating the matter dynamics). Our results suggests that this likely to be more subtle than hitherto anticipated; this issue is worth investigating further. Similar issues would arise if we were to use the causal holographic information which is also reduced relative to the vacuum.

The causal wedge analysis was primarily aimed at exhibiting an explicit example of a causally trivial spacetime where $\cwedge$ has non-trivial topology (leading to the causal information surface $\csf{\cal A}$ being disconnected). The analysis in fact follows similar lines as described in \cite{Hubeny:2013gba} and we were able to see that the causal wedge topology is non-trivial only when the geometry admits null circular orbits as conjectured there.

However, a non-trivial corollary of our investigation was the fact that for small regions the casual wedges reach deeper in to the bulk than they do in the vacuum \AdS{} spacetime when we have condensation of modes with alternate boundary condition. While this potentially indicates a conflict with the time-delay theorems in asymptotically AdS spacetimes \cite{Woolgar:1994ar,Gao:2000ga} we argued that one has no real tension. Said differently, the bulk geometries with $\vev{{\cal O}_1}\neq 0$ satisfy the basic consistency requirement of 
bulk causality being commensurate with boundary causality. We conjecture that the bulk-cone singularities inherited in the the dual CFT states always lie inside the boundary light-cone.\footnote{ Strictly speaking our argument was phrased by examining the radial null geodesics. In order to argue that there are no bulk-cone singularities we need to show that the bulk geodesics carrying non-trivial angular momentum along $\partial_\varphi$ also do not travel faster through the bulk.}

The most interesting example of a spacetime where the bulk causal structure is incommensurate with the boundary casual structure is provided by the singular soliton geometry discussed in Appendix~\ref{app:singular}. Here the null geodesics traveling through the bulk allow for faster communication between boundary points than geodesics localised on the boundary, i.e., we encounter bulk-cone singularities outside the boundary light-cone. This acausal behaviour of boundary correlation functions is intimately tied to the time-like singularity in the bulk spacetime. As we have remarked this is qualitatively similar to the behaviour of bulk-cone singularities expected in the negative mass Schwarzschild-\AdS{d+1} black hole (despite the fact that in the present case the singular soliton has positive ADM mass). This feature alone should be sufficient to rule out the geometry as being dual to a sensible field theory state. We would like to argue that the criterion for admissible singularities in the bulk geometry dual to a quantum field theory should be enlarged from those discussed in \cite{Horowitz:1995ta, Gubser:2000nd} to include as an explicit criterion the compatibility of the bulk and boundary causal structures. One could indeed use this argument to rule out the negative mass Schwarzschild-AdS solution being dual to a sensible CFT state (this is \emph{a priori} independent of the rationale presented in \cite{Horowitz:1995ta} to rule out these geometries), but the operative point is that is also works for seemingly reasonable geometries such as \eqref{eq:DGglobalsingular}.

We also computed the causal holographic information contained in polar-cap regions in these states. To our knowledge this is the first time this quantity has been calculated in gravitational solutions with non-trivial bulk matter fields. Our results are in accordance with the properties discussed in \cite{Hubeny:2012wa,Wall:2012uf,Hubeny:2013gba}.  It would be interesting to compute this quantity when $\theta_{\cal A} > \theta_{\cal A}^*$ and track how it changes through the break-up of the causal information surface. We anticipate that $\chi_{\cal A} (\theta_{\cal A})$ would be continous but not differentiable as we go through the transition from a single surface to a disconnected one (the argument is similar to the one for entanglement entropy given in \cite{Hubeny:2013gta}).

Finally, we should note that all of our discussion has focused on states where single-trace operators ${\cal O}_{\Delta_\pm}$ get vacuum expectation values. It is interesting to ask what happens when we have states where the scalar satisfies multi-trace boundary conditions. These types of geometries have been discussed earlier in the context of designer gravity \cite{Hertog:2004ns} and this dial was also exploited in  \cite{Gentle:2011kv}  to explore the enlarged space of solutions. If the scalar satisfies a multi-trace boundary condition then generically it would not be true that vacuum AdS would be a ground state of the system (it could be a designer gravity soliton). It would be interesting to examine how the multi-trace boundary conditions affect the entanglement content of the state --- we expect there to be some non-trivial interplay owing once again to the slow fall-off of the scalar field

\subsection*{Acknowledgements}

It is a pleasure to thank Gary Horowitz, Veronika Hubeny, Per Kraus, Henry Maxfield,  Rob Myers, Tadashi Takayanagi and Benjamin Withers for helpful discussions. We like to thank Centro de Ciencias de Benasque Pedro Pascual and LMU Munich for their kind hospitality during the course of the project. MR in addition acknowledges the hospitality of KITP (Santa Barbara) during the concluding stages of the project.  SAG is supported by an STFC studentship, an STFC STEP award and by NSF grant PHY-13-13986. MR is supported in part by the  STFC Consolidated Grant ST/J000426/1.

\appendix

\section{Perturbative solitons}
\label{app:pertubative}

Here we consider solitons that can be constructed in a perturbative expansion around AdS$_4$.  For a given choice of quantization, we choose the expectation value $\langle\mathcal{O}_{\Delta}\rangle$ to be our small parameter.  A suitable ansatz, as well as other asymptotic data as functions of this parameter, can be found in Appendix B of  \cite{Gentle:2011kv}. Here we present the results for $g(r)$, which we will need in \S\ref{sec:ee}.

Perturbing around global AdS$_4$ with $\vev{{\cal O}_{\Delta}}=\varepsilon$ for general $\Delta=\frac{3}{2} \pm \sqrt{\frac{9}{4} + m_\phi^2}$ we find
\begin{equation}
g(r) =r^2+1- \frac{\Delta (2 \Delta-3 )}{6} \,   r^2  \, _2F_1\left(\frac{3}{2},\Delta ;\frac{5}{2};-r^2\right)\, \varepsilon^2 + O(\varepsilon^4)
\end{equation}
to lowest non-trivial order in $\varepsilon$. For $\Delta=2$ this reduces to
\begin{equation}
g(r) = r^2+1+\frac{1}{2} \left(\frac{1}{r^2+1}-\frac{\tan ^{-1}r}{r}\right) \varepsilon^2 +O(\varepsilon^4)
\end{equation}
whereas for $\Delta=1$ we have
\begin{equation}
g(r) = r^2+1+\frac{ \left(r-\tan ^{-1}r\right)}{2 r}\, \varepsilon^2 +O(\varepsilon^4)
\end{equation}

\section{Singular soliton}
\label{app:singular}

The $U(1)^4$ truncation admits an analytical solution carrying a non-trivial scalar profile, parametrized by the scalar fall-off $\phi_1$:
\begin{equation}\label{eq:DGglobalsingular}
\begin{gathered}
g(r)  = r^2 + 1+\frac{\phi_1^2}{2}+\frac{\phi_1^2}{2 r^2},\quad e^{-\beta(r)} g(r)=  r^2+1, \\
\quad A_t(r)=0, \quad \phi(r) = \sqrt{2}\sinh^{-1} {\frac{\phi_1}{\sqrt{2}r}} 
\end{gathered}
\end{equation}
These solutions are neutral solitons with no conserved charges that are instead supported by a non-trivial scalar field.   These solitons correspond to (2+1)-dimensional CFTs in which the  $\Delta =1$ operator ${\cal O}_{1}$ spontaneously acquires an expectation value $\langle\mathcal{O}_1\rangle=\phi_1$ that breaks the $U(1)$ global symmetry on the boundary. From the asymptotic scalar fall-off it is clear that there is no deformation in the dual CFT. Crucially, this family of solutions is singular because the Ricci scalar diverges at $r=0$.  The planar limit of this solution is the planar limit of the $\Delta=1$ connected unbounded branch of regular solitons in this theory. We find it interesting to examine the role played by the singularity in the context of our analysis.

\begin{figure}[h!]
\vskip1em
\begin{center}
\includegraphics[width=0.35\textwidth]{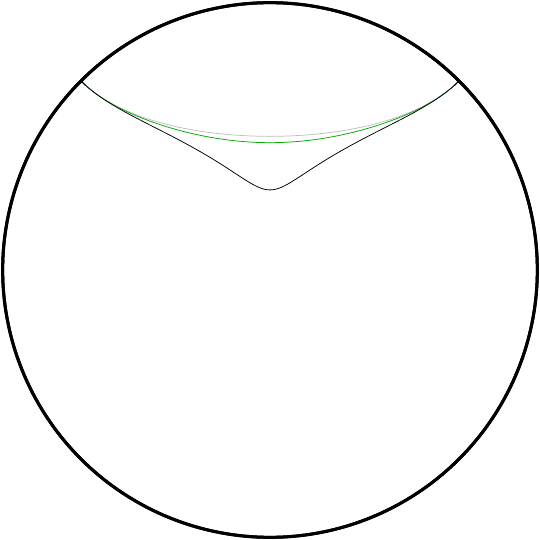}
\setlength{\unitlength}{0.1\columnwidth}
\end{center}
\vskip-0.5em
\caption{Minimal surfaces in $\vev{{\cal O}_1}$ soliton geometries in the $U(1)^4$ truncation.  The regular soliton (green) and the singular soliton (black) both have $\phi_1=1.28(4)$ and global AdS$_4$ is shown in grey for comparison.  All are anchored at $\theta_{\cal A}=\frac{\pi}{4}$.}
\label{fig:DGsurface}
\end{figure}

As described in \S\ref{sec:alternate}, in the alternate quantization we expect minimal surfaces $\extr{\cal A}$ to penetrate deeper into the spacetime than for AdS$_4$. The presence of the singularity however exacerbates this phenomenon, as illustrated in Fig.~\ref{fig:DGsurface}.

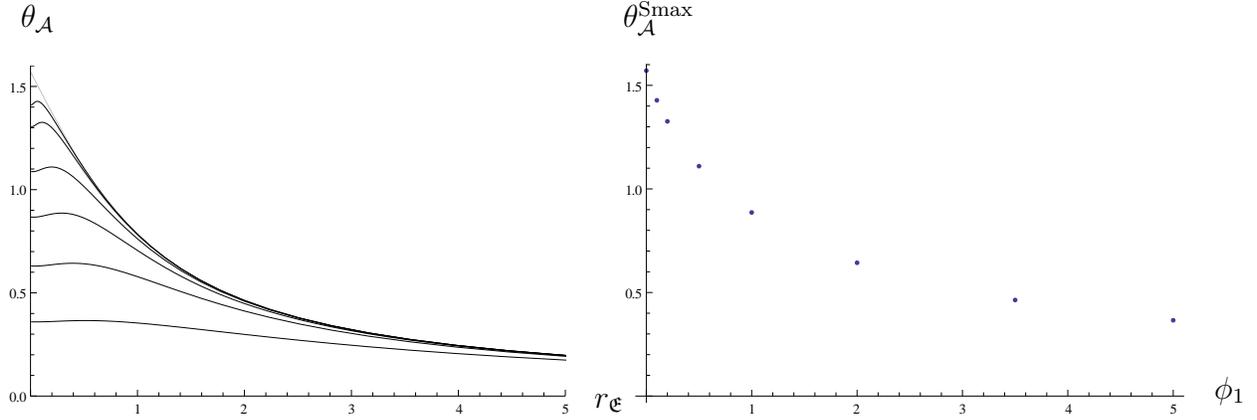
\begin{figure}[h!]
\vskip1.5em
\begin{center}
\includegraphics[width=0.45\textwidth]{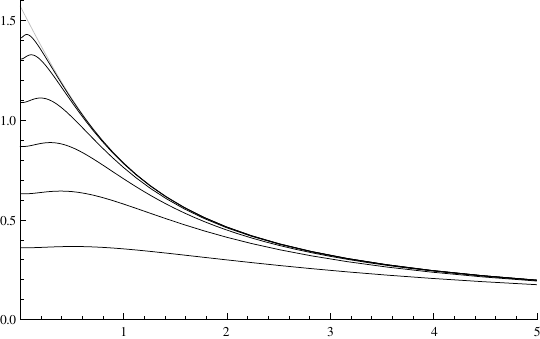}
\hskip1.5em
\includegraphics[width=0.45\textwidth]{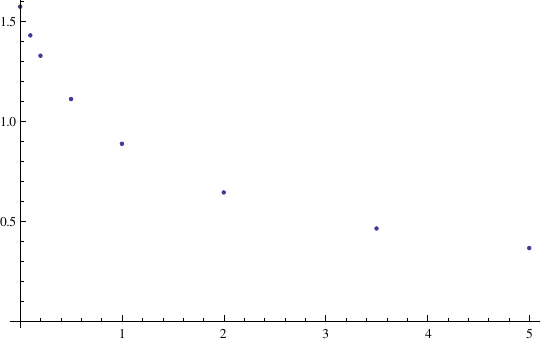}
\setlength{\unitlength}{0.1\columnwidth}
\begin{picture}(0.3,0.4)(0,0)
\put(-9.4,3.2){\makebox(0,0){$\theta_{\mathcal{A}}$}}
\put(-4.8,0.1){\makebox(0,0){$\rE$}}
\put(-4.4,3.2){\makebox(0,0){$\theta_{\mathcal{A}}^{\mathrm{Smax}}$}}
\put(0.2,0.2){\makebox(0,0){$\phi_1$}}
\end{picture}
\end{center}
\vskip-0.5em
\caption{Left: Families of minimal surfaces in different singular soliton geometries. From top (AdS$_4$) to bottom: $\phi_1=\color{light-gray}{0}\color{black}{,0.1,0.2,0.5,1,2,5}$.  Right: $\theta_{\mathcal{A}}^{\mathrm{Smax}}$ as a function of $\phi_1$.}\label{fig:DGFixedPhi1Set}
\end{figure}

A second curious feature is that not all choices of boundary region result in a smooth bulk minimal surface. In fact, for fixed $\phi_1$ there exists a critical $\theta_{\cal A}^{\mathrm{Smax}}$ above which there are no smooth minimal area surfaces. One can see this by following our algorithm of finding minimal surfaces with smooth initial conditions at $\rE$; for  $\theta_{\cal A} < \theta_A^{\mathrm{Smax}}$ we can integrate out to the boundary and find the appropriate region ${\cal A}$. The behaviour of the curve $\theta_{\cal A}(\rE)$ is shown in Fig.~\ref{fig:DGFixedPhi1Set}; it is non-monotone with a characteristic maximum $\theta_{\cal A}^{\mathrm{Smax}}$ for a given value of $\phi_1$. This in particular means that for larger regions there is no smooth solution to the minimal surface problem. We also display $\theta_{\cal A}^{\mathrm{Smax}}(\phi_1)$ in Fig.~\ref{fig:DGFixedPhi1Set} to show that the region where smooth minimal surfaces exists gets smaller as we crank up the scalar expectation value.

\begin{figure}[!htb]
\begin{center}
\includegraphics[width=0.35\textwidth]{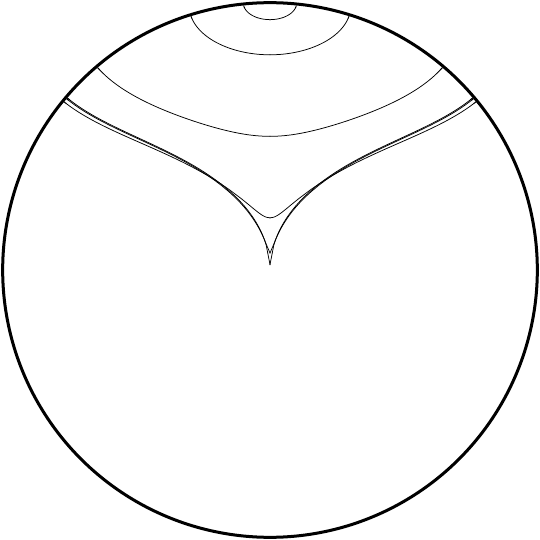}
\hskip2em
\includegraphics[width=0.35\textwidth]{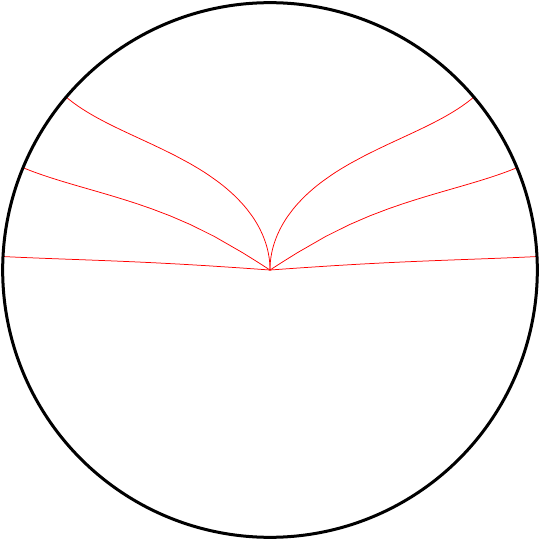}
\end{center}
\caption{Minimal surfaces in the singular soliton with $\phi_1=1$. Left: smooth surfaces with $\rE=10^x$ for $x=1,0.5,0,-0.5,-1,-1.5$ from top to bottom.  Right: cuspy surfaces with $\theta_0=0.01,1,1.5$ from top to bottom.}
\label{fig:AllSingularSolitonSurfaces}
\end{figure}

The only surfaces that exist for $\theta_{\cal A}> \theta_{\cal A}^{\mathrm{Smax}}$ are singular minimal surfaces that are anchored at the timelike singularity at $r=0$.\footnote{ Strictly speaking the symmetric surface $\extr{\cal A}^\sharp$, i.e., $\theta(r) = \frac{\pi}{2}$, is itself singular since it passes though the singular point $r=0$.}
We can allow these surfaces to have a cusp at $r=0$. Since we no longer require smoothness, radial parametrisation is straightforward to implement and $\theta(r)$ has the following expansion for small $r$:
\begin{equation}
\theta(r)= \theta_0 + \frac{ \cot \theta_0}{3 \phi_1^2}\, r^2 -\frac{ \cot \theta_0 \left(7 \cot ^2\theta_0+36 \phi_1^2+81\right)}{270 \phi_1^4}\, r^4 + O(r^6)
\end{equation}
We plot examples of smooth and cuspy minimal surfaces in Fig.~\ref{fig:AllSingularSolitonSurfaces}.

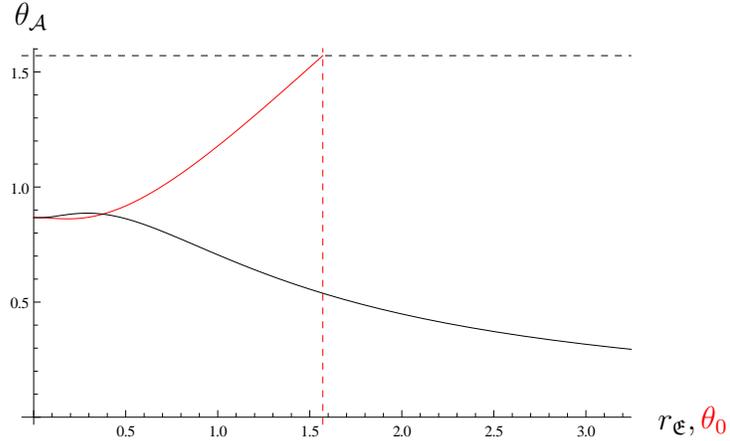
\begin{figure}[h!]
\vskip1.5em
\begin{center}
\includegraphics[width=0.5\textwidth]{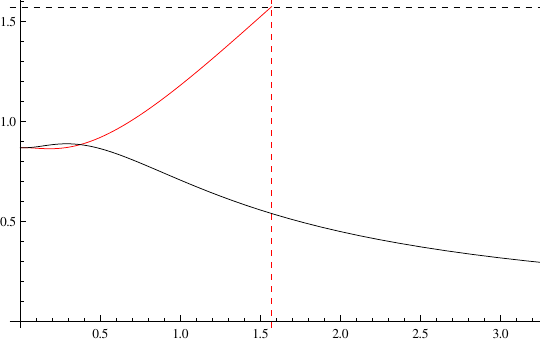}
\setlength{\unitlength}{0.1\columnwidth}
\begin{picture}(0.3,0.4)(0,0)
\put(-5,3.4){\makebox(0,0){$\theta_{\mathcal{A}}$}}
\put(0.2,0.1){\makebox(0,0){$\rE$,}}
\put(0.5,0.15){\makebox(0,0){$\color{red}{\theta_0}$}}
\end{picture}
\end{center}
\vskip-0.5em
\caption{Families of minimal surfaces in the singular soliton with $\phi_1=1$. The black curve is for smooth surfaces and the red curve is for surfaces with a cusp at the singularity.  The horizontal and vertical dashed lines are at $\theta_{\mathcal{A}}=\frac{\pi}{2}$ and $\theta_0=\frac{\pi}{2}$, respectively.}\label{fig:DGSmoothVsCusp}
\end{figure}

To get a full picture of the set of surfaces in the geometry, in  Fig.~\ref{fig:DGSmoothVsCusp} we illustrate the smooth and singular minimal surfaces available for a given $\theta_{\cal A}$ paremeterised by $\rE$ and $\theta_0$ respectively. We note that the non-monotone behaviour is quite pronounced with characteristic oscillations near $\rE=0$ or $\theta_0=0$. Not only do we find that the family of smooth surfaces connects to those with a cusp at some $\theta_{\mathcal{A}}^{\mathrm{crit}}$, but it also appears that there can be between one and five surfaces with the same value of $\theta_{\mathcal{A}}$.  Away from an `overlap region' in $\theta_{\mathcal{A}}$ there is only one. 
As a result of these curious features, the singular soliton geometry (even excising $r=0$ from the spacetime manifold) cannot be foliated by \emph{smooth} extremal surfaces, in contrast to the regular solitons of the $U(1)^4$ truncation. 

Just as for other $\Delta=1$ solitons, the entanglement entropy for the singular soliton is lower than that for the AdS$_4$ vacuum.  In particular, for the symmetric surface we have
\begin{align}\label{eq:singarea}
\frac{1}{4\pi\,c_\text{eff}} \, \Delta S_{\cal A}^\sharp &= 2\pi \int_0^{\infty} dr\, \frac{r}{\sqrt{r^2+1}}\left(\frac{1}{\sqrt{1+\phi_1^2/(2 r^2)}}-1\right)\nonumber\\
&= 2\pi\left(1-E\left(1-\frac{\phi_1^2}{2}\right)\right) \leq 0
\end{align}
where $E(x)$ is the complete elliptic integral of the second kind with the property $E(1)=1$.  This regulated area decreases monotonically from zero as a function of $\phi_1$.  

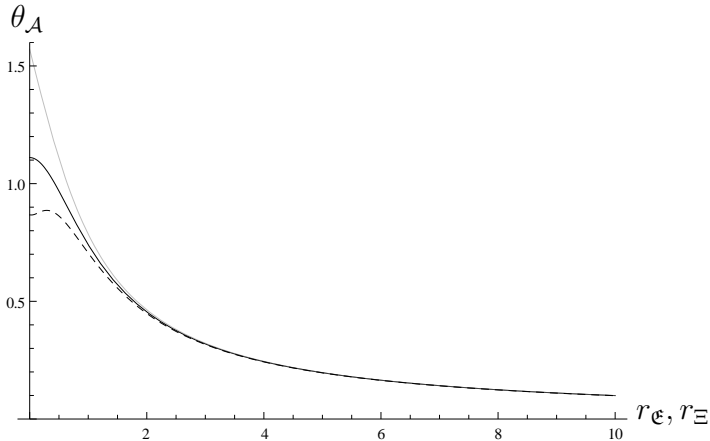
\begin{figure}[h!]
\vskip1em
\begin{center}
\includegraphics[width=0.5\columnwidth]{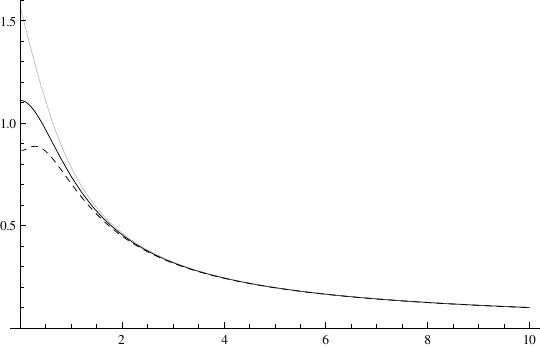}
\setlength{\unitlength}{0.1\columnwidth}
\begin{picture}(0.3,0.4)(0,0)
\put(-5.,3.4){\makebox(0,0){$\theta_{\cal A}$}}
\put(0.2,0.2){\makebox(0,0){$\rE, \rX$}}
\end{picture}
\end{center}
\vskip-0.5em
\caption{Deepest extent of $\csf{\cal A}$ (solid line) and $\extr{\cal A}$ (dashed line) surfaces in the singular soliton with $\phi_1=1$ and AdS$_4$ shown in grey for comparison.}
\label{fig:SingularrEandrX}
\end{figure}

The singular solitons \eqref{eq:DGglobalsingular} also have several curious causal properties. Solving \eqref{eq:wedgedepth} explicitly we can find the radial extent of the causal information surface analytically:
\begin{equation}
\rX=\sqrt{\left(1-\frac{\phi_1^2}{2}\right)\csc^2\left(\sqrt{1-\frac{\phi_1^2}{2}}\,\theta_{\cal A} \right)-1}
\end{equation}
This has the correct AdS$_4$ limit of $\cot\theta_{\cal A}$ for $\phi_1=0$. In Fig.~\ref{fig:SingularrEandrX} we compare this result to a family of smooth minimal surfaces in the same geometry.
Note that we always find $\rE<\rX<r_{\mathrm{AdS}}$ at fixed $\theta_{\cal A}$ for the singular soliton --- a property that only held for small $\theta_{\mathcal{A}}$ in regular solitons in the same theory.  

A natural consequence of the fact that the casual information surface penetrates further into the bulk irrespective of the size of the region is that the geometry allows faster communication through the bulk than across the boundary. Explicitly, the time of flight through the bulk between anti-podal points on the boundary is given by
\begin{equation}
\Delta t = 2 \int_{0}^{\infty} \frac{dr}{(r^2+1)\sqrt{1+\frac{\phi_1^2}{2r^2}}} = \frac{2 \cos ^{-1}\frac{\phi_1}{\sqrt{2}}}{\sqrt{1-\frac{\phi_1^2}{2}}} \in [0,\pi] 
\end{equation}
Thus, the time of flight in this geometry is bounded from \emph{above} by the AdS result and can be made arbitrarily small by increasing $\phi_1$.  This means that correlation functions in the dual state have additional Lorentzian bulk-cone singularities \cite{Hubeny:2006yu} \emph{outside} the light cone.  Said differently, one can transmit signals in this state faster than the speed of light. 

Naively this result seems to contradict the time-delay theorem of \cite{Gao:2000ga}. \emph{A priori} the Einstein-Maxwell-scalar Lagrangian \eqref{eq:generalaction} obeys the null energy condition and the solution is appropriately asymptotically AdS (despite the slow fall-off of the scalar field). However, the theorem of \cite{Gao:2000ga} assumes that the spacetimes in question are smooth, which \eqref{eq:DGglobalsingular} is clearly not. So \emph{per se}, there is no obvious conflict with the general expectations. We believe the situation is analogous to that of the negative mass Schwarzschild-AdS black hole where too one can engineer faster communication through the bulk than across the boundary.




\providecommand{\href}[2]{#2}\begingroup\raggedright\endgroup

\end{document}